\documentclass[%
 reprint,
superscriptaddress,
prl,
 amsmath,amssymb,
 aps,
]{revtex4-2}
\usepackage{graphicx}
\usepackage{dcolumn}
\usepackage{bm}
\usepackage{xcolor}
\usepackage{physics, amsmath}
\usepackage{hyperref}
\usepackage[percent]{overpic}
\usepackage{bbm}

\usepackage{enumitem}

\usepackage[many]{tcolorbox}
\usetikzlibrary{calc}

\definecolor{myblue}{RGB}{1,103,153}

\tcbset{mystyle/.style={
  breakable,
  enhanced,
  outer arc=0pt,
  arc=0pt,
  size = fbox,
  colframe=myblue,
  colback=myblue!10,
  attach boxed title to top left,
  boxed title style={
    colback=myblue,
    outer arc=0pt,
    arc=0pt,
    top=1pt,
    bottom=1pt,
    },
  fonttitle=\sffamily
  }
}

\newtcolorbox{TextBox}[1]{
  mystyle,
}

\begin{document}

\preprint{APS/123-QED}

\title{Quantum vs Classical Erasure: 
Equal Bounds but Unequal Costs}

\author{Jan Neuser}
    \email{jan.neuser.physics@gmail.com}
    \affiliation{Vienna Center for Quantum Science and Technology, Atominstitut, TU Wien, 1020 Vienna, Austria}
 \author{Jake Xuereb}
   \affiliation{Vienna Center for Quantum Science and Technology, Atominstitut, TU Wien, 1020 Vienna, Austria}
 \author{Pharnam Bakhshinezhad}
    \affiliation{Vienna Center for Quantum Science and Technology, Atominstitut, TU Wien, 1020 Vienna, Austria}
\author{Marcus Huber}
    \affiliation{Vienna Center for Quantum Science and Technology, Atominstitut, TU Wien, 1020 Vienna, Austria}
    \affiliation{Institute for Quantum Optics and Quantum Information - IQOQI Vienna, Austrian Academy of Sciences, Boltzmanngasse 3, 1090 Vienna, Austria}
    
\date{\today}

\begin{abstract}
Irreversibility has a fundamental thermodynamic cost, erasing information inevitably generates heat. This connection is quantified by the Landauer bound, which gives the minimum dissipation needed to erase a single bit of information. While this bound applies in both classical and quantum settings, it is saturated only in idealised limits of infinite resources. Here, we provide a unified first principles description of finite-resource erasure in both classical and quantum systems. We begin by proving the communal folklore that in the idealised regime the erasure cost of a bit encoded in a quantum or classical system is the same. Despite this, we show that their practical implementation differs substantially: achieving comparable erasure quality in quantum systems requires more control, larger accessible energy gaps and longer operation times. Classical protocols can achieve the erasure of a comparable quantum protocol under far weaker constraints which we expose in trade-off relations.
Our results explain why practical erasure schemes fall short of Landauer’s bound and show that classical systems enjoy several fundamental thermodynamic advantages.
\end{abstract}

\maketitle
\noindent
\textbf{Introduction.} 
Landauer's bound on dissipation establishes a fundamental link between information processing and physics. It states that erasing one bit of information requires dissipating at least $k_\text{B}T\log(2)$ into a thermal environment at temperature $T$ \cite{Original_Landauer, bennettThermodynamicsComputationReview1982a}.
However Landauer's bound is an idealization and only reachable in an infinite time, energy or complexity limit whereas finite-time erasure carries with it significantly higher costs \cite{Taranto_2023,Proesmans_2020, Rolandi_2023, VanVu2025_TimeCostError}. Today the best chips in classical computation are still orders of magnitude above this bound \cite{ho2023limitsenergyefficiencycmos, Pop2010}, while an accounting of the thermodynamics of quantum computers is an active endeavour~\cite{carrascocodina2026energyefficiencyquantumcomputers,auffeves,meier2025autonomousquantumprocessingunit}. 

The difference between erasing bits of information encoded in quantum and classical systems with access to finite resources is of both fundamental and practical interest~\cite{Reeb_and_Wolf,
buffoniCooperativeQuantumInformation2023,tarantoExponentialImprovementQuantum2020,tarantoEfficientlyCoolingQuantum2025,xuerebCoolingQubitUsing2025,millerQuantumFluctuationsHinder2020,dagoFoolingLandauerBound2026}. Classically, bits are encoded in many-body systems with a large number of (quantum) subsystems per logical value, providing intrinsic robustness to microscopic noise but suffering larger entropy and energy scales \cite{Plenio_2001, Koomey2011_efficiency}. In contrast, quantum bits (qubits) are microscopic and have far lower energy scales, which may suggest an intrinsic efficiency advantage. While Landauer's bound sets a universal lower limit, realistic erasure protocols operate far from this ideal. This raises a central question which we address in this work: \textit{how does the physics of quantum and classical systems encoding a bit of information impact erasure}?

\begin{figure*}
\centering
\begin{minipage}{0.49\textwidth}
\llap{\hspace{2mm}\textbf{(a)}\hspace{-0.2cm}}
\begin{overpic}[width=\linewidth, trim={0.1cm 0cm 0cm 0cm},clip]{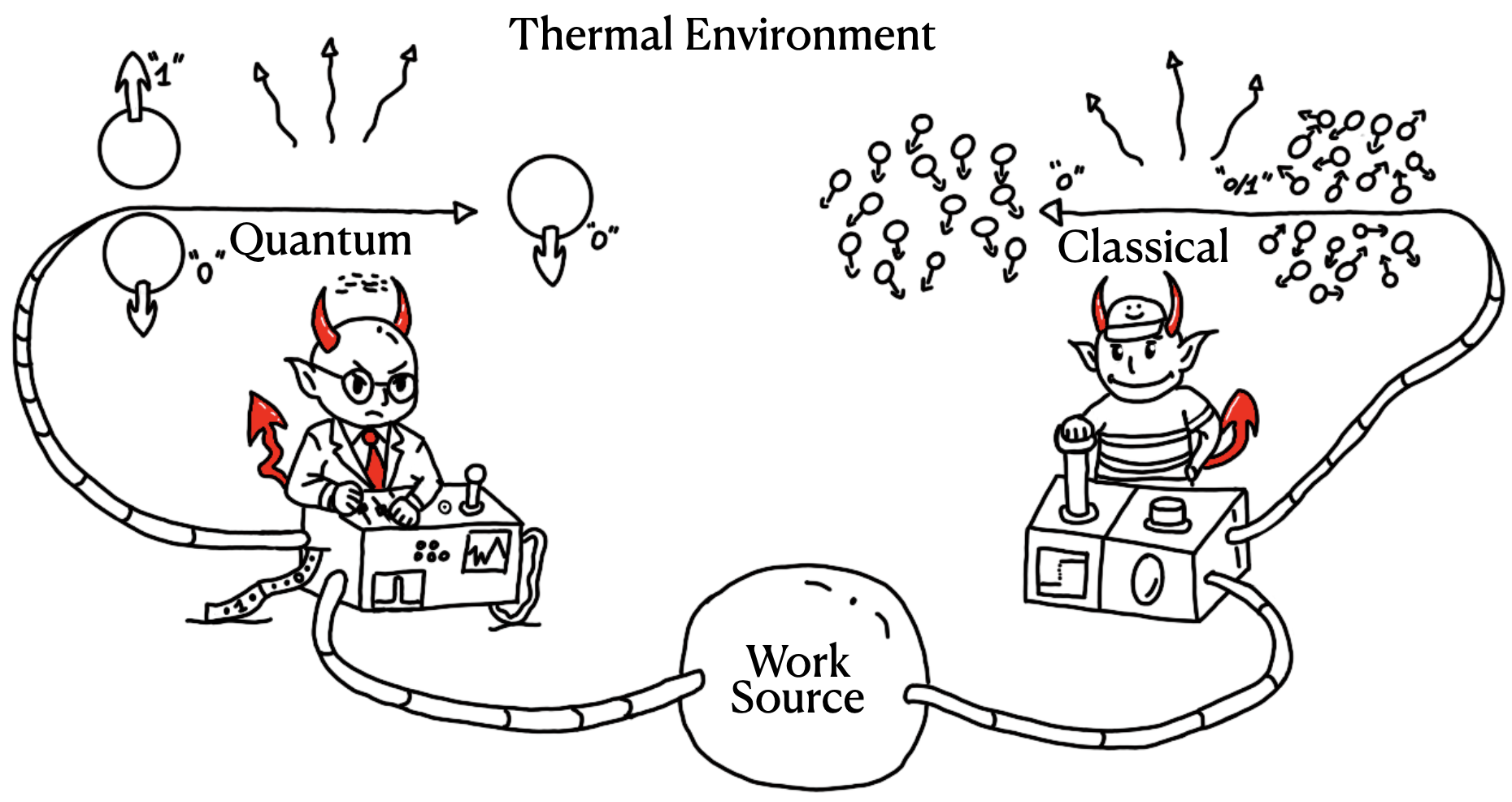}
\end{overpic}
\end{minipage}
\hfill
\begin{minipage}{0.49\textwidth}
\llap{\hspace{2mm}\textbf{(b)}\hspace{-0.2cm}}
\begin{overpic}[width=\linewidth]{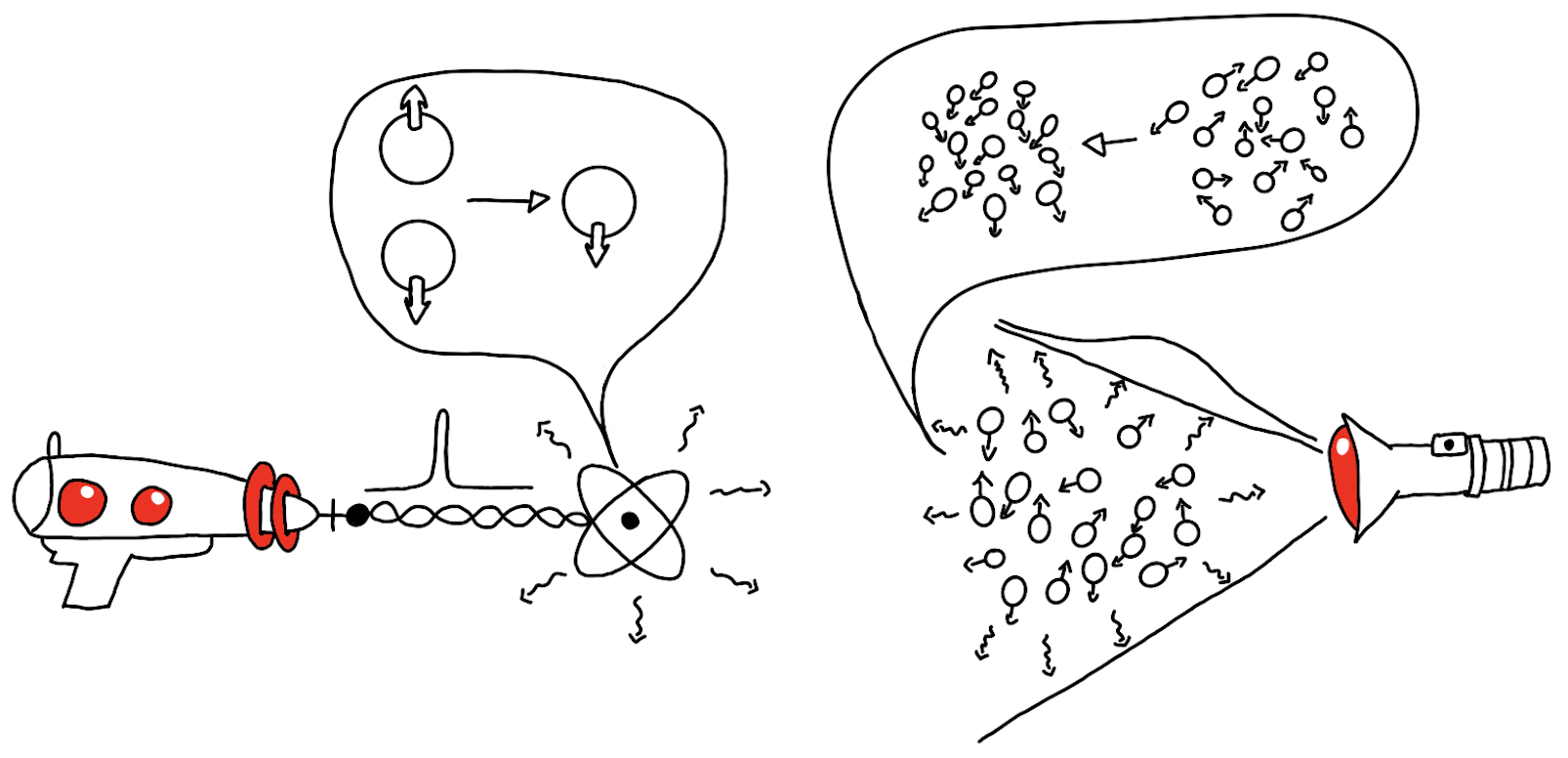}
\end{overpic}
\end{minipage}

\caption{\textbf{(a)} General setup we consider: the systems are embedded in a thermal environment, both the classical and quantum control use a work source to perform the respective erasure protocol. \textbf{(b)} Even though they use the same resources the control requirements differ: the quantum bit requires a sharp control frequency, while the classical bit allows for a wider spectrum.}
\label{fig: combined sketches}
\end{figure*}

We find that without restrictions on control or time, classical and quantum protocols require comparable total resources, this is irrespective of the energy scale of the system. Relaxing these idealised assumptions, we show that in realistic settings erasing a bit encoded in a quantum system imposes stricter operational requirements than erasing a classical bit, precise control, long interaction times and large energy gaps are needed. We then examine a trade-off between these requirements by comparing a quantum bit erased through a finite number of cooling interactions to a classical bit erased through a single cooling interaction acting on a macroscopic number of subsystems. We show that a classically encoded bit can always be erased at a lower dissipative cost in a single round, simply by increasing the number of subsystems that encode it. Finally, we find that this redundancy also makes the classical bit inherently more robust, timing errors directly degrade the fidelity of quantum erasure, but a large subsystem number suppresses the same errors exponentially in the classical bit.

Whilst previous work has explored the cost of erasing a bit of information in a quantum system in the presence of coherence~\cite{millerQuantumFluctuationsHinder2020, VanVu2022_FiniteTime} with limitations to finite time~\cite{modi_ft,Proesmans_2020} and using collective cooling mechanisms~\cite{buffoniCooperativeQuantumInformation2023, Lipka2025_QuadraticScalingLB, Aimet2025_ManybodyLB}, our work represents the first attempt at a comparative analysis of the thermodynamics of classical and quantum bit erasure under finite resources. In particular we do so by analysing both erasure scenarios using the same fully self-contained framework~\cite{brunner_12, Mitchison_2019}, allowing for a fair comparison.

\textbf{Classical and quantum bits.} To address the question of erasure cost in both quantum and classical systems, we first need to introduce what we understand by a classical and a quantum bit (qubit). For the qubit information is encoded in a $2$-dimensional Hilbert space $\text{span}\{|0\rangle, |1\rangle\}$~\cite{Nielsen2010}.    
Classical bits on the other hand are encoded into a vast state space of a many body system. Here only the logical 0 and 1 states implemented via the coarse-graining of the $d$-dimensional space need to be distinguishable. The logical subspaces are defined by projectors $\Pi_0$ and $\Pi_1 = \mathbbm{1}_d-\Pi_0$. The choice of $\Pi_0$ has to be reasonable, as both states are equally likely to appear in processing and should be equally easy to populate and address. So we will always choose $d/2$ states to define the 1 and the other $d/2$ to define the 0 subspace.

To make a fair comparison, we require that initially both system are in a state without any knowledge about the encoding, implying initially
\begin{align}
\rho_{\text Q} &= \frac{\mathbbm{1}_2}{2}, &
\rho_{\text C} &= \frac{\mathbbm{1}_d}{d} \,  .
\label{eq: MM state of both}
\end{align}
giving $\bra{0}\rho_{\text Q}\ket{0} = \text{tr}\{\Pi_0 \rho_{\text C}\} = 1/2$.  We then define the fidelity of the erasure protocol $\mathcal{F}$ by the total population in the logical 0 subspace i.e. to erase to a fidelity of $\mathcal{F}$ we require that $\rho_{\text Q} \rightarrow \rho_{\text Q}'$ such that $\bra{0}\rho_{\text Q}'\ket{0} = \mathcal{F}$ and similarly for the classical system $\rho_{\text C} \rightarrow \rho'_{\text C}$ with $\text{tr}\{\Pi_0 \rho'_{\text C}\} = \mathcal{F}$, perfect erasure is defined by $\mathcal{F} = 1$.

\textbf{Equivalent Landauer Bounds.} 
We can now answer our first question: \emph{is the fundamental limit on dissipation due to erasure the same for our definitions of classical and quantum bits?} By the Clausius inequality $\beta\Delta Q_{\text B}\geq -\Delta S_S$, where $\Delta S_S$ is the entropy change due to the erasure. Erasing a qubit to fidelity $\mathcal{F}$ thus requires us to dissipate at least $ \beta \Delta Q_{\text B}^{\text (Q)} \geq  \log(2)+\mathcal{F} \log (\mathcal{F}) +(1-\mathcal{F})  \log (1-\mathcal{F})$ of heat into the environment at inverse temperature $\beta$, see Eq.~\eqref{eq: finite landauer bound}. For a $d$-level system the entropy change is generally larger implying $\beta \Delta Q_{\text B}^{\text (Q)} \leq\beta \Delta Q_{\text B}^{\text (C)} $, but coincides with that of the qubit when the populations are uniformly distributed within each logical subspace, more details are given in the supplementary material~\cite{supmat}.

This observation carries an important implication for our comparison the \textit{optimal} erasure protocol for a bit encoded in a $d$-level system, coincides with the protocol for a two-level system. But a $d$-level system can host more information and so generically undergoes a larger entropy change upon erasure, implying a larger lower bound. But by uniformly distribution of the target populations one recovers $\Delta S^{(d)}_S = \Delta S^{(2)}_S$ and thus the two Landauer bounds coincide, but this equality is a property of the bound alone, it offers only limited insight into the physics of an achievable erasure protocol.

\textbf{Comparing Asymptotic Classical \& Quantum Erasure Protocols.} Despite this equivalence, it is unclear whether these two scenarios will be identical when considering a 
realistic erasure protocol. To ensure a thermodynamically complete analysis, we analyse the erasure protocol using energy-preserving unitaries so that the energetics of the protocol are all accounted for. Such protocols decrease the entropy of the target system, thus throughout the manuscript the terms cooling and erasure will be used interchangeably. To cool a system we need to couple it to at least two out-of-equilibrium resources (in the spirit of the absorption refrigerator model~\cite{skrzypczyk_2010,brunner_12,Mitchison_2019}). 
To be precise, the choice of two out-of-equilibrium resources is needed to go beyond simple thermalization, allowing for genuine erasure beyond the temperature of the coldest reservoir.

The energy conservation of the time evolution $U,$ implies $[U, H_\text{tot}] = 0$, where $H_\text{tot} = H_{\text S} + H_{\text B}+ H_{\text W}$ is the total Hamiltonian, consisting of system, thermal bath/environment and a work source, with $H_A=\sum_k E_k^A\ketbra{E_k^A}{E_k^A}$ for $A\in\{{\text S,B,W}\}$.

The work source is left general, but for realism we restrict it to interact with the system around a single frequency $\Delta E_\text{W}$, which could be a single laser frequency for sideband or Doppler cooling \cite{Yan2018_SingleAtomLP, Leibfried2003_RMP}, a magnetic field applied to a collection of spins or a specific bias voltage.

Initially the total state will be decoupled s.t. $\rho_\text{tot} = \rho_{\text Q, C} \otimes \tau_{\beta} \otimes \rho_{\text W}$,
where $\rho_{\text Q,C}$ is given by Eqs.~\eqref{eq: MM state of both}, $ \tau_{\beta}$ is the Gibbs state of the thermal bath at inverse temperature $\beta$, i.e. $\tau_\beta = e^{- \beta H_{\text B}}/\tr \{ e^{- \beta H_{\text B}}\}$ and $\rho_{\text W}$ of the initial state of the work source, which needs to be an out of equilibrium state w.r.t $\tau_\beta$. To erase energy preserving the system beyond the temperature of the thermal bath we need to pick an energy gap of the system $E^{\text S}_i-E^{\text S}_{i'}:=\Delta E_{\text S}^{\text Q,C}$, where typically $\ket{E^{\text S}_i}$ would belong to the logical 1 and $\ket{E^{\text S}_{i'}}$ to the 0 subspace. Further we pick energy gaps of the bath $E^{\text B}_i-E^{\text B}_{i'}:=\Delta E_{\text B}$ and work source $E^{\text W}_i-E^{\text W}_{i'}:=\Delta E_{\text W}$ respectively, such that 
\begin{equation}
  \Delta E_{\text S}^{\text Q,C}=  \Delta E_{\text B}-\Delta E_{\text W}\,.
  \label{eq: energy resonance condition}
\end{equation}
For simplicity, let us denote these bath and work source eigenstates by $H_{\text B/W}|0/1\rangle=E^{\text B/W}_{i/i'}|0/1\rangle$.  With this energetic structure, the system gap $\Delta E_{\text S}^{\text Q,C}$ may be cooled using only one generator (ie. one control frequency) represented by the set of energy preserving interaction Hamiltonians
\begin{equation}
\label{eq: General interaction Hamiltonian}
    H_{j} = g \left(\sigma_{\text S} \otimes \sigma^\dagger_{\text{B},j} \otimes \sigma_{\text{W},j} + \text{h.c.} \right)  \, , 
    %
    %
\end{equation}
with $g$ the interaction strength and $\sigma_{A} = \ketbra{E_{i'}^{A}}{E_i^{A}}$, with $A\in\{{\text S,B,W}\}$, is a lowering operator of the respective system gaps, note that due to Eq.~\eqref{eq: energy resonance condition} this will be energy preserving. We now restrict the control to a single frequency, i.e. $\Delta E_{\text W}$ is the same for all system gaps. This could for example be a single laser frequency, a single magnetic field, or any other control field characterised by one dominant frequency, as opposed to a tunable or multi-frequency source. Further in Eq.~\eqref{eq: General interaction Hamiltonian} the index $j$ labels distinct bath and work source constituents that this single frequency can address, e.g. via successive or parallel interactions. Note that the thermal bath and work source are assumed to be macroscopic, i.e. their states are unchanged by dissipation due to erasure.

\begin{figure*}
\centering
\begin{minipage}{0.49\textwidth}
\llap{\hspace{-1cm}\textbf{(a)}\hspace{-1.2cm}}
\begin{overpic}[width=\linewidth, trim={0.0cm 0cm 0cm 0cm},clip]{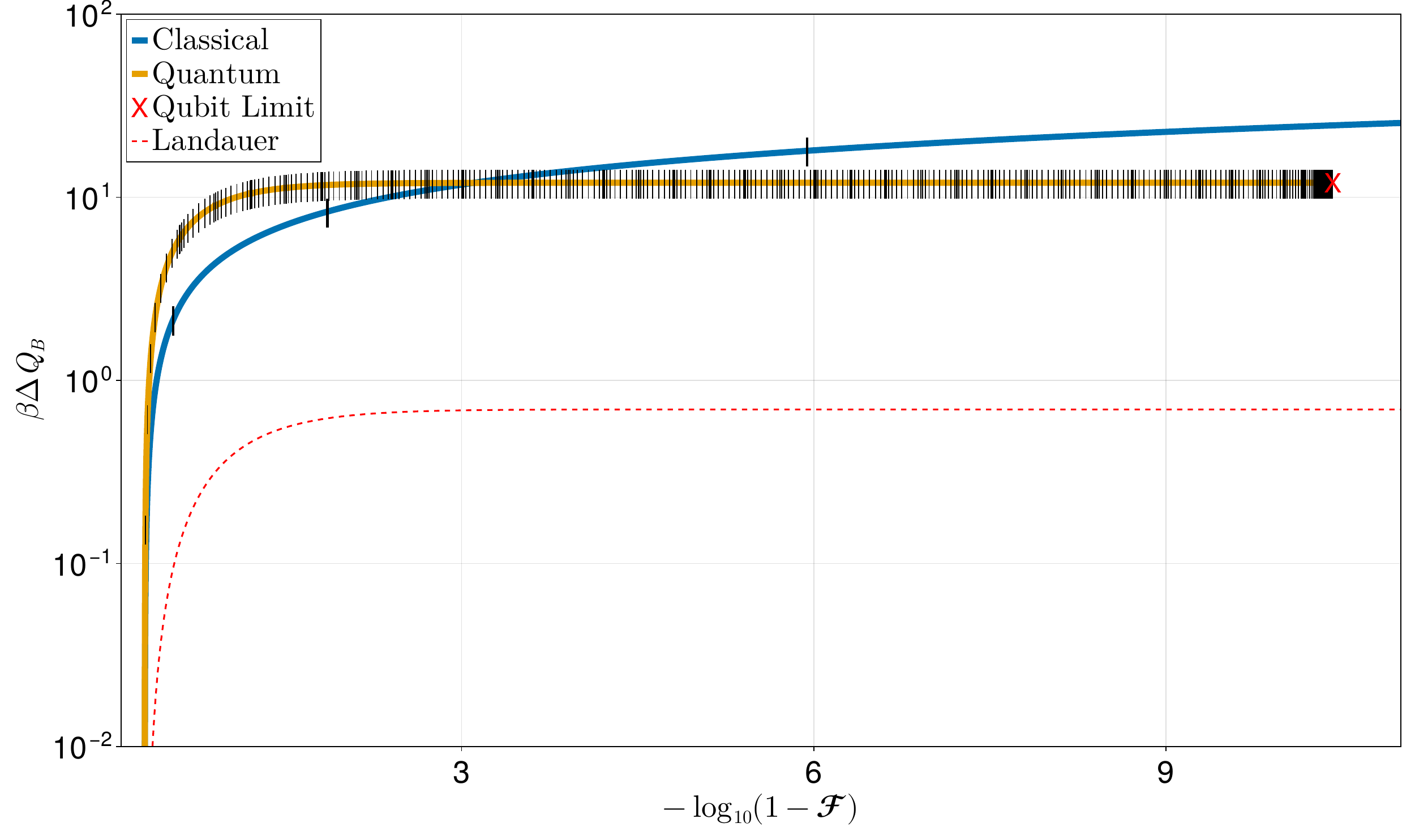}
\end{overpic}
\end{minipage}
\hfill
\begin{minipage}{0.49\textwidth}
\llap{\hspace{-1cm}\textbf{(b)}\hspace{-1.2cm}}
\begin{overpic}[width=\linewidth, trim={0.0cm 0cm 0cm 0cm},clip]{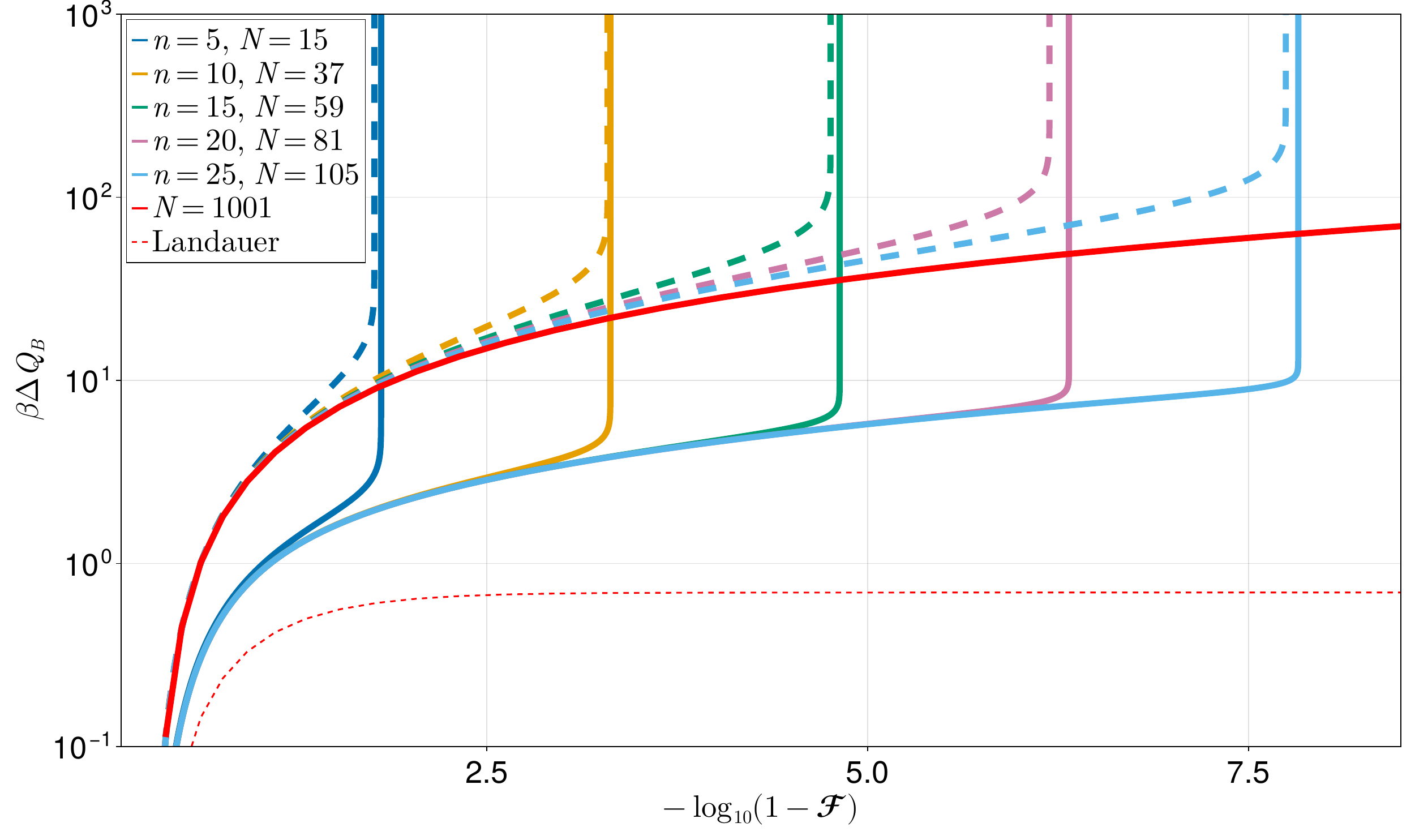}
\end{overpic}
\end{minipage}
\caption{\textbf{(a)} Dissipation versus erasure fidelity, shown as $-\log_{10}(1-\mathcal{F})$, for quantum (orange) and classical (blue) erasure. The classical bit consists of $N=10^{11}+1$ qubits,
the time between black bins is kept constant and the red cross marks the infinite time limit of the qubit. The work qubits are at a temperature $T_{\text W} = 1000\text{K}$, in the quantum case the thermal qubits use a gap of $\Delta E_{\text B} = 10 \omega$ and for the classical bit $\Delta E_{\text B} = 10^{-5} \omega$, with $\omega/(2\pi \hbar ) =  5 \text{GHz}$ .
\textbf{(b)} Dissipation for quantum erasure after $n$ perfectly timed swaps (solid) and classical erasure using a single parallelized interaction of $N$ qubits (dashed). For each $n$, $N$ is the largest odd integer satisfying $N\leq N_{\min}(n)$. The control gap $\Delta E_{\text B}$ is varied to approach the maximal fidelity $\mathcal{F}^{\max}$. Red curve shows a classical bit with $N=1001$. In both panels, $\Delta E_{\text S}^{\text Q,C}/(2\pi\hbar)=\omega$ and $T_{\text B}=0.1\text{K}$, the red dashed curve indicate the finite-erasure Landauer bound of Eq.~\eqref{eq: finite landauer bound}.}
\label{fig: combined Plots}
\end{figure*}

This interaction, induces a (partial) swap between the system and a \emph{virtual qubit}~\cite{Baumer2019imperfect,Xuereb_2023}. We first look at an idealized asymptotic setting i.e. infinitely many perfectly timed sequential cooling interactions, where at each step we interact with a refreshed thermal bath- and work source system, which leads to~\cite{supmat}
\begin{equation}
    p_i \rightarrow p_i^* = \frac{p_i+ p_{i'}}{1+e^{-\beta\Delta E_{\text B}} \mathcal{W} } \, , 
     \label{eq: infintie swap limit p_i}
\end{equation}
where $p_i$ and $p_{i'}$ are the populations of $\ket{E^{\text S}_{i}}$ and $\ket{E^{\text S}_{i'}}$, $\mathcal{W} = \frac{\langle0|\rho_{\text W}|0 \rangle}{\langle 1|\rho_{\text W}|1\rangle} = \frac{w_0}{w_1}$ and  $w_1$ is the population of $|E_{i}^{\text W}\rangle$ and $w_0$ of $|E_{i'}^{\text W}\rangle$.

We introduce the swap fidelity $F = p_i^*/(p_i + p_{i'})$, with $F\in [0,1] $, which lets us know the fraction/percentage of the total available population we have moved to the 0 space. Note that for the qubit the swap fidelity and the erasure fidelity coincide $F = \mathcal{F}$, as $p_i + p_{i'} = 1$ and $p_{i'}^* = \mathcal{F}$ is the final 0 subspace population, but for the classical bit this is generally not the case. In this asymptotic limit we find the dissipation into the heat bath is given by $\Delta Q_{\text B} = \Delta E_\text{B} \Delta p$, where $\Delta p= p_i^*-p_i$, more details in the supplementary~\cite{supmat}, which leads to
\begin{equation}
    \Delta Q_{\text B} = \frac{(F-1) p_i+F p_{i'} }{\beta} \log\left( \frac{F}{1-F}\mathcal{W}\right)\, ,
    \label{eq: rescaled dissipation by fidelity} 
\end{equation}
In the qubit case $p_i=p_{i'}=1/2$ as the levels we pick to swap with are simply the logical states $\ket1$ and $\ket0$. Now the dissipation obtained by Eq.~\eqref{eq: rescaled dissipation by fidelity} diverges as $F\rightarrow1$, i.e. perfect erasure requires an infinite heat dissipation into the thermal reservoir, in agreement with the Nernst unattainability principle \cite{Levy_2012, Masanes2017,Taranto_2023}.

Next assume the classical system is a $d$-dimensional system with a specific structure, namely we can pair the $d$ energy levels into $d/2$ pairs, each pair consisting of one state from the logical 0 subspace and one from the logical 1 subspace, such that every pair shares the same gap $\Delta E_\text{S}^\text{C}$. As all $d/2$ pairs have an identical gap, a single work source frequency $\Delta E_{\text W}$ suffices to address every pair simultaneously, exactly as required by our single control frequency assumption. Given access to infinitely many such perfectly timed swaps, we show in the supplemental material that the resulting dissipation coincides with that of the qubit i.e. $\Delta Q^{\text (C)}_{\text B} = \Delta Q^{\text (Q)}_{\text B}$, see~\cite{supmat}. In other words, even for single control frequencies, the idealised cost of erasure for quantum systems and classical systems (coarse grained from equally spaced d-level systems addressable by a single frequency) remains the same.

\textbf{A Bit Encoded in a Collection of Subsystems.}{\textemdash} 
Instead of a single $d$-dimensional system, we now turn to a more realistic description of a classical bit as a many-body system. We consider a classical bit encoded in many weakly interacting quantum subsystems, which admits an effective mean-field description~\cite{Fiorelli_2023}. Such weakly interacting subsystems are naturally addressed collectively, so that a single driving frequency can act on the entire ensemble. This setting is natural, as for example hard drives manipulate spins in magnetic domains via external magnetic fields~\cite{1996_Magnetism_Leccabue_F}, while solid state drives control electron transport through bias voltages applied to floating gates~\cite{Taur_Ning_2009}.

The logical $0/1$ value is then determined by a majority vote over the constituents, for example the sign of the total magnetization or the charge stored on a floating gate. Each constituent contributes to this vote via its own local 0 and 1 subspace, for a spin, spin up could contribute to local 1 and spin down to local 0. But constituents could also be higher-dimensional, in which case we partition them, assigning several microstates to each local outcome. If these microstates can be paired across the local 0 and 1 subspaces by a single gap, a single control frequency addresses all pairs simultaneously and the constituent, however high-dimensional, reduce to an effective qubit. We prove this explicitly in the end matter, which allows us to restrict the following analysis to qubit constituents of the classical bit.

Accordingly, the erasure fidelity is the population in the majority-$0$ subspace, with
\begin{equation}
    \Pi_0 = \sum_{s \in \mathcal{S}} \ketbra{s}{s} \, , 
\end{equation}
where $\mathcal{S}$ contains all length $N$ bit strings with more $0$s than $1$s. We take $N$ to be odd so that the logical $0$ and $1$ subspaces contain the same number of microstates.

This construction differs from the exemplary $d$-level system considered above. There, we chose the target populations to be distributed uniformly within each logical subspace. Here, instead, we cool each of the $N$ constituent two-level systems identically and individually using the same tripartite cooling interaction E~ This leaves the populations over the $d = 2^N$ microstates non-uniformly distributed within each logical subspace giving a suboptimal change in entropy. The asymptotic bound derived in the appendix~\cite{supmat} remains a lower bound for the majority-vote encoding, as illustrated in Fig.~\ref{fig: combined Plots}(a,b). Whilst this appears to be a step backwards, we show in the following that precisely this structure yields several advantages once we restrict the protocol to finite resources.

\textbf{The Role of Time.}{\textemdash}We now examine the role of time and control in this more realistic description of a classical bit. To erase a qubit, we employ Hamiltonians ${H_j}$ of the form of Eq.~\eqref{eq: General interaction Hamiltonian}, with $\sigma_{\text S}=\ketbra{0}{1}$ and each $H_j$ coupling the qubit to a distinct virtual qubit. Applying the corresponding unitaries $U_i=e^{-iH_i\tau_{\text opt}}$ sequentially for a time $\tau_{\text opt}= \frac{\pi}{2g}$, results in 
\begin{equation}
    U_{\text Q} =  \prod_{i = 1}^n U_i\, , 
\end{equation}
which requires a total time of $n \tau_{\text opt}$. Alternatively we can apply all operations simultaneously by introducing a hermitian $H$, such that $e^{-iH\tau'} = U_{\text{Q}}$ for some finite time $\tau'$. While formally possible, this requires a $(2n+1)$-partite interaction, which substantially increases the control complexity. One thus faces a trade-off between simple control over long times and complicated control over short times \cite{Taranto_2023,Taranto_2025}.

This is in stark contrast to the classical system, as here each $H_j$ addresses a different subsystem or a transition of a higher-dimensional subsystem, through a corresponding operator $\sigma_{\text{S},j}$. Consequently, $[H_i,H_j]=0$ for $i\neq j$, as the system operators commute, while the bath and work operators act on distinct refreshed ancillas and therefore also commute. If we now restrict to a single cooling step per effective subsystem i.e. each $H_j$ is applies only once, we find
\begin{equation}\label{eq: classical unitary}
    U_{\text C} = \prod_{i = 1}^N e^{-i  H_i \tau_{\text opt}} = e^{-i \sum_i H_i \tau_{\text opt}} \, , 
\end{equation}
which requires just time $\tau_{\text opt}$, plus we are only using tripartite interactions. But in this case, each subsystem interacts with the virtual-qubit subspace only once and is therefore cooled less than the qubit. This reduced cooling per subsystem is not detrimental, as shown below and could in any case be compensated by applying $U_{\text C}$ repeatedly, at the cost of a longer interaction time~\cite{supmat}.

Figure~\ref{fig: combined Plots}(a) compares the resulting finite-time protocols. The qubit requires substantially longer interaction times to reach fidelities comparable to those of the classical. Remarkably the dissipation is comparable even though the classical encoding considered contains $11$ orders of magnitude more energy, the majority voting converts weak, parallel cooling of many subsystems into fast and accurate erasure at comparable dissipation.

As we care about the finite-time regime, it is natural to ask: \emph{how does the dissipation in the classical and quantum settings depend on the protocol duration?} 
Assuming a protocol time of $n \tau_{\text opt}$ the qubit is limited to a fidelity of
\begin{equation}
    \mathcal{F}_{\text Q}^{\text{max}}(n) = 1-\frac{(1-w_1)^n}{2} \, , 
    \label{eq: Divpoint qubit}
\end{equation}
which can only be reached in the limit of infinite dissipation $\Delta Q_{\text B} \rightarrow \infty$, as detailed in the supplementary material~\cite{supmat}. Where we assume the work source purity $1/\mathcal{W}$ to be independent of the gap $\Delta E_{\text B}$. Thus the qubit is subject to a trade-off between dissipation and protocol time; increasing one allows for a decrease of the other~\cite{supmat}.

The classical system faces a different trade-off, as all $N$ subsystems are addressed in parallel, erasure requires only the minimal interaction time regardless of $N$. But the achievable fidelity can still be increased arbitrarily simply by enlarging the subsystem number $N$~\cite{supmat}.

This prompts the question; \textit{if a qubit is allowed $n$ perfectly timed swaps, how many effective two-level subsystems must a classical bit have in order to achieve the same erasure in a single interaction?} To investigate this let us assume $\mathcal{W}$ to be independent of $\Delta E_B$ and give both the classical and the quantum bit the same bath temperature $\beta$ and gaps $\Delta E_{\text B}$ and $\Delta E_{\text S}^{\text Q,C}$. For large $N$ and $n$ we find $N_{\text min} \propto n $ see Eq.~\eqref{eq: minimum N for same divergence point} for details. $N_{\text min}$ is the minimum number of subsystems the classical system needs to be comprised of in order to achieve the same asymptotic fidelity $\mathcal{F}_{\text Q}^{\text{max}}(n) = \mathcal{F}_{\text C}^{\text{max}}(N_{\text min})$ (reached only at infinite dissipation $\Delta Q_B\rightarrow \infty$) . Thus if $N > N_{\text min}$ the classical system will be able to reach fidelities that are inaccessible to the qubit in a protocol of total time $n\tau_{\text opt}$, we display this behaviour in Fig.~\ref{fig: combined Plots} (b).

Finite time qualitatively distinguishes the two encodings. For any fixed number of interactions, the qubit encounters a maximal fidelity that can only be approached at divergent dissipation. The classical bit instead trades control time for subsystem number, parallel interactions allow it to reach high-fidelities in minimal time.

\textbf{Finite Temperature Work Source.}{\textemdash}
Finally, let's address the work source. So far, we assumed that the populations $w_0$ and $w_1$ and consequently $\mathcal{W}$ are independent of the bath gap $\Delta E_{\text B}$. However, if the work source is described by thermal populations at inverse temperature $\beta_{\text W}$, this corresponds to the infinite-temperature limit $\beta_{\text W} = 0$, as the Gibbs ratio defines $\mathcal{W}$ by Eq.~\eqref{eq: energy resonance condition}.

For $\beta_{\text W} \neq 0$ he target qubit after $n$ interactions with the thermal work source reaches a fidelity of
\begin{align}
\mathcal{F}_\text{Q} 
&= \frac{1-q^n}{1 +e^{-\beta_{\text W} \Delta E_{\text S}^Q}e^{-(\beta- \beta_{\text W})\Delta E_{\text B} }} + \frac{q^n}{2}, 
\label{eq: F_Q hot work source}
\end{align}

where $q = (1+ e^{-\beta_{\text W} (\Delta E_{\text B}-\Delta E_{\text S}^Q) -\beta \Delta E_{\text B}})/(\mathcal{Z}_{\text B} \mathcal{Z}_\text{W})$ and $\mathcal{Z}_{\text B} = \tr\{e^{- \beta H_{\text B}}\}$, $\mathcal{Z}_\text{W} = \tr\{e^{- \beta_{\text W} H_{\text W}}\}$ are partition functions of the hot/work source and cold bath respectively.
Each population exchange brings the system closer to thermalization to the virtual qubit temperature~\cite{Clivaz_2019}
$
    \beta_{\text V} = \frac{(\beta- \beta_\text{W} ) \Delta E_{\text B} }{\Delta E_{\text S}^Q} +\beta_\text{W}\, , 
$
which depends on the temperature difference $\beta-\beta_{\text W}$ and thereby the temperature difference limits the achievable fidelity. Compensating for this finite-temperature effect requires a larger bath gap $\Delta E_{\text B}$, which in turn increases the associated dissipation as $\Delta Q_{\text B}^\text{Q} = \Delta E_{\text B} (\mathcal{F}- \frac{1}{2})$ \cite{supmat,Taranto_2025}.

As in the finite-time setting, the classical encoding can instead compensate for imperfect cooling by increasing the number of constituent subsystems~\cite{supmat}. A larger $N$ reduces the population bias required in each subsystem. So we can surmise that imperfect sources of work hardly affect classical erasure but exacerbate the limitations faced by quantum erasure.

\textbf{Robustness of the protocol.}{\textemdash} Having examined finite resource constraints, we now consider imperfect control. As a paradigmatic example, we consider timing errors, which could come from imperfect knowledge of the coupling strengths $g$ to the virtual subspace or an imprecise clock \cite{janovitch2025activequantumreservoirengineering, Xuereb_2023}. We model this by letting the interaction duration fluctuate as $\tau = \tau_{\text opt} + \varepsilon/g$, with $\varepsilon \sim \mathcal{N}(0,s^2)$ a normal distribution with mean 0 and variance $s^2$. In the Supplemental Material, we show that the resulting error is exponentially suppressed in subsystem number $N$ for the classical bit see Eq.~\eqref{eq: Exponential decrease classical for mistimings}. On the contrary it directly degrades the reachable qubit fidelity see Eq.~\eqref{eq: reachable p0 qubit mistimings}. The classical bit therefore tolerates a wide spread of mistimed, or equivalently off-resonant, control frequencies and still erases with high fidelity, majority voting suppresses the resulting errors, but the qubit requires a genuinely sharp control frequency. We capture this need for sharp control in Fig.~\ref{fig: combined sketches}(b).

\textbf{Conclusion.}{\textemdash}
In this work, we have investigated the thermodynamic cost of erasing information encoded in classical and quantum systems under finite-resource constraints. Although Landauer's principle assigns the same asymptotic cost to both tasks, our results show that their practical realization can differ dramatically. The key distinction lies not in the lower bound but in the resources required to approach it.

By comparing a classically encoded bit, defined by microstates of many constituents, with a quantum bit stored in a single two-level system, we find that classical information erasure can achieve arbitrarily high fidelity through the collective of many moderately erased subsystems in finite time. In contrast, quantum erasure has substantially more demanding requirements on energy scales, time and control. We also stress that while our setting may appear fairly idealized, the end matter provides further clarification and shows how our assumptions map onto realistic physical scenarios.

Our findings therefore reveal a clear separation between classical and quantum information erasure. More broadly, these results highlight that the thermodynamic challenge of quantum information processing arises not from a different fundamental limit, but from the difficulty of approaching the same limit under realistic constraints.

\textbf{Acknowledgments.}{\textemdash} JN, JX and MH acknowledge support from the European Research Council (ERC Project ’Cocoquest’ 101043705). P.B. acknowledges that financial support was provided by the Austrian Science Fund (FWF) through the StandAlone grant P 35810-N and P 36633-N. We acknowledge Carlos Pineda, Carlos Viviescas, Niyusha Hosseini, Nayeli Rodríguez, Caroline Käse and Alberto Rodriguez for initial fruitful discussions. JN acknowledges Jakob Huber for the explanation of the working principle of a MOS transistor and Joelle Broch for feedback on the manuscript. 

\textbf{Tool Disclosure}{\textemdash} We used Mathematica 15.0 and Chat GPT for symbolic analysis. The LLM was also used to assist with literature review and text clarification. The central idea, derivations, figures and writing of the manuscript is due to the authors in its entirety. The code to generate the plots is available on \href{https://github.com/neusiboi/Quantum-vs-Classical-Erasure}{Github}.

\bibliography{References}

@article{Leibfried2003_RMP,
  title = {Quantum dynamics of single trapped ions},
  author = {Leibfried, D. and Blatt, R. and Monroe, C. and Wineland, D.},
  journal = {Rev. Mod. Phys.},
  volume = {75},
  issue = {1},
  pages = {281--324},
  numpages = {0},
  year = {2003},
  month = {Mar},
  publisher = {American Physical Society},
  url = {https://link.aps.org/doi/10.1103/RevModPhys.75.281}
}

@article{Aimet2025_ManybodyLB,
	author = {Aimet, Stefan and Tajik, Mohammadamin and Tournaire, Gabrielle and Sch{\"u}ttelkopf, Philipp and Sabino, Jo{\~a}o and Sotiriadis, Spyros and Guarnieri, Giacomo and Schmiedmayer, J{\"o}rg and Eisert, Jens},
	doi = {10.1038/s41567-025-02930-9},
	isbn = {1745-2481},
	journal = {Nature Physics},
	number = {8},
	pages = {1326--1331},
	title = {Experimentally probing Landauer's principle in the quantum many-body regime},
	url = {https://doi.org/10.1038/s41567-025-02930-9},
	volume = {21},
	year = {2025}}

@article{Lipka2025_QuadraticScalingLB,
  title = {Minimizing Dissipation via Interacting Environments: Quadratic Convergence to Landauer Bound},
  author = {Lipka-Bartosik, Patryk and Perarnau-Llobet, Mart\'{\i}},
  journal = {Phys. Rev. Lett.},
  volume = {135},
  issue = {17},
  pages = {170404},
  numpages = {7},
  year = {2025},
  month = {Oct},
  publisher = {American Physical Society},
  doi = {10.1103/5w1r-1nzs},
  url = {https://link.aps.org/doi/10.1103/5w1r-1nzs}
}

@ARTICLE{Koomey2011_efficiency,
  author={Koomey, Jonathan and Berard, Stephen and Sanchez, Marla and Wong, Henry},
  journal={IEEE Annals of the History of Computing}, 
  title={Implications of Historical Trends in the Electrical Efficiency of Computing}, 
  year={2011},
  volume={33},
  number={3},
  pages={46-54},
  doi={10.1109/MAHC.2010.28}}

@article{Plenio_2001,
   title={The physics of forgetting: Landauer’s erasure principle and information theory},
   volume={42},
   ISSN={1366-5812},
   url={http://dx.doi.org/10.1080/00107510010018916},
   DOI={10.1080/00107510010018916},
   number={1},
   journal={Contemporary Physics},
   publisher={Informa UK Limited},
   author={Plenio, M. B. and Vitelli, V.},
   year={2001},
   month=Jan, pages={25–60} }

@article{VanVu2022_FiniteTime,
   title={Finite-Time Quantum Landauer Principle and Quantum Coherence},
   volume={128},
   ISSN={1079-7114},
   url={http://dx.doi.org/10.1103/PhysRevLett.128.010602},
   number={1},
   journal={Physical Review Letters},
   publisher={American Physical Society (APS)},
   author={Van Vu, Tan and Saito, Keiji},
   year={2022},
   month=Jan }

@article{VanVu2025_TimeCostError,
  title = {Time-Cost-Error Trade-Off Relation in Thermodynamics: The Third Law and Beyond},
  author = {Van Vu, Tan and Saito, Keiji},
  journal = {Phys. Rev. X},
  volume = {15},
  issue = {4},
  pages = {041029},
  numpages = {26},
  year = {2025},
  month = {Nov},
  publisher = {American Physical Society},
  doi = {10.1103/l6b9-rg1j},
  url = {https://link.aps.org/doi/10.1103/l6b9-rg1j}
}

@article{Yan2018_SingleAtomLP,
  title = {Single-Atom Demonstration of the Quantum Landauer Principle},
  author = {Yan, L. L. and Xiong, T. P. and Rehan, K. and Zhou, F. and Liang, D. F. and Chen, L. and Zhang, J. Q. and Yang, W. L. and Ma, Z. H. and Feng, M.},
  journal = {Phys. Rev. Lett.},
  volume = {120},
  issue = {21},
  pages = {210601},
  numpages = {6},
  year = {2018},
  month = {May},
  publisher = {American Physical Society},
  doi = {10.1103/PhysRevLett.120.210601},
  url = {https://link.aps.org/doi/10.1103/PhysRevLett.120.210601}
}

@misc{supmat,
  note = {In the supplemental material we provide the derivation of the finite-error Landauer bound, the virtual-qubit cooling dynamics, the equality in dissipation of a d-level system under the pairing condition and the qubit, the majority vote scaling in subsystem number, the minimum subsystem number for equal reachable fidelity and the effects of imperfect timing.}}

@article{Original_Landauer,
  author={Landauer, R.},
  journal={IBM Journal of Research and Development}, 
  title={Irreversibility and Heat Generation in the Computing Process}, 
  year={1961},
  volume={5},
  number={3},
  pages={183-191},
  doi={10.1147/rd.53.0183}}

@Book{Nielsen2010,
  author    = {Nielsen, Michael A. and Chuang, Isaac L.},
  publisher = {Cambridge University Press},
  title     = {Quantum {C}omputation and {Q}uantum {I}nformation},
  year      = {2010},
  edition   = {10},
  doi       = {10.1017/CBO9780511976667},
  place     = {Cambridge},
}

@article{Baumer2019imperfect,
  doi = {10.22331/q-2019-06-24-153},
  url = {https://doi.org/10.22331/q-2019-06-24-153},
  title = {Imperfect {T}hermalizations {A}llow for {O}ptimal {T}hermodynamic {P}rocesses},
  author = {B{\"{a}}umer, Elisa and Perarnau-Llobet, Mart{\'{i}} and Kammerlander, Philipp and Wilming, Henrik and Renner, Renato},
  journal = {{Quantum}},
  issn = {2521-327X},
  publisher = {{Verein zur F{\"{o}}rderung des Open Access Publizierens in den Quantenwissenschaften}},
  volume = {3},
  pages = {153},
  month = jun,
  year = {2019}
}

@article{Masanes2017,
  author       = {Masanes, Lluis and Oppenheim, Jonathan},
  title        = {A general derivation and quantification of the third law of thermodynamics},
  journal      = {Nature Communications},
  volume       = {8},
  pages        = {14538},
  year         = {2017},
  doi          = {10.1038/ncomms14538},
  url          = {https://doi.org/10.1038/ncomms14538}
}

@article{skrzypczyk_2010,
  title = {How Small Can Thermal Machines Be? The Smallest Possible Refrigerator},
  author = {Linden, Noah and Popescu, Sandu and Skrzypczyk, Paul},
  journal = {Phys. Rev. Lett.},
  volume = {105},
  issue = {13},
  pages = {130401},
  numpages = {4},
  year = {2010},
  month = {Sep},
  publisher = {American Physical Society},
  doi = {10.1103/PhysRevLett.105.130401},
  url = {https://link.aps.org/doi/10.1103/PhysRevLett.105.130401}
}

@article{auffeves,
  title = {Optimizing Resource Efficiencies for Scalable Full-Stack Quantum Computers},
  author = {Fellous-Asiani, Marco and Chai, Jing Hao and Thonnart, Yvain and Ng, Hui Khoon and Whitney, Robert S. and Auff\`eves, Alexia},
  journal = {PRX Quantum},
  volume = {4},
  issue = {4},
  pages = {040319},
  numpages = {40},
  year = {2023},
  month = {Oct},
  publisher = {American Physical Society},
  doi = {10.1103/PRXQuantum.4.040319},
  url = {https://link.aps.org/doi/10.1103/PRXQuantum.4.040319}
}

@article{Clivaz_2019,
   title={Unifying Paradigms of Quantum Refrigeration: A Universal and Attainable Bound on Cooling},
   volume={123},
   ISSN={1079-7114},
   url={http://dx.doi.org/10.1103/PhysRevLett.123.170605},
   number={17},
   journal={Physical Review Letters},
   publisher={American Physical Society (APS)},
   author={Clivaz, Fabien and Silva, Ralph and Haack, Géraldine and Brask, Jonatan Bohr and Brunner, Nicolas and Huber, Marcus},
   year={2019},
   month=oct }

@article{Reeb_and_Wolf,
   title={An improved Landauer principle with finite-size corrections},
   volume={16},
   ISSN={1367-2630},
   url={http://dx.doi.org/10.1088/1367-2630/16/10/103011},
   DOI={10.1088/1367-2630/16/10/103011},
   number={10},
   journal={New Journal of Physics},
   publisher={IOP Publishing},
   author={Reeb, David and Wolf, Michael M},
   year={2014},
   month=oct, pages={103011} }

@article{Taranto_2023,
    title = {Landauer Versus Nernst: What is the True Cost of Cooling a Quantum System?},
  author = {Taranto, Philip and Bakhshinezhad, Faraj and Bluhm, Andreas and Silva, Ralph and Friis, Nicolai and Lock, Maximilian P.E. and Vitagliano, Giuseppe and Binder, Felix C. and Debarba, Tiago and Schwarzhans, Emanuel and Clivaz, Fabien and Huber, Marcus},
  journal = {PRX Quantum},
  volume = {4},
  issue = {1},
  pages = {010332},
  numpages = {61},
  year = {2023},
  month = {Mar},
  publisher = {American Physical Society},
  url = {https://link.aps.org/doi/10.1103/PRXQuantum.4.010332}
}

@article{Rolandi_2023,
   title={Finite-time Landauer principle beyond weak coupling},
   volume={7},
   ISSN={2521-327X},
   url={http://dx.doi.org/10.22331/q-2023-11-03-1161},
   DOI={10.22331/q-2023-11-03-1161},
   journal={Quantum},
   publisher={Verein zur Forderung des Open Access Publizierens in den Quantenwissenschaften},
   author={Rolandi, Alberto and Perarnau-Llobet, Martí},
   year={2023},
   month=nov, pages={1161} }

@article{Taranto_2025,
   title={Efficiently Cooling Quantum Systems with Finite Resources: Insights from Thermodynamic Geometry},
   volume={134},
   ISSN={1079-7114},
   url={http://dx.doi.org/10.1103/PhysRevLett.134.070401},
   number={7},
   journal={Physical Review Letters},
   publisher={American Physical Society (APS)},
   author={Taranto, Philip and Lipka-Bartosik, Patryk and Rodríguez-Briones, Nayeli A. and Perarnau-Llobet, Martí and Friis, Nicolai and Huber, Marcus and Bakhshinezhad, Pharnam},
   year={2025},
   month=feb }

@article{Levy_2012,
   title={Quantum refrigerators and the third law of thermodynamics},
   volume={85},
   ISSN={1550-2376},
   url={http://dx.doi.org/10.1103/PhysRevE.85.061126},
   number={6},
   journal={Physical Review E},
   publisher={American Physical Society (APS)},
   author={Levy, Amikam and Alicki, Robert and Kosloff, Ronnie},
   year={2012},
   month=jun }

@article{modi_ft,
  title = {Universal Bound on Energy Cost of Bit Reset in Finite Time},
  author = {Zhen, Yi-Zheng and Egloff, Dario and Modi, Kavan and Dahlsten, Oscar},
  journal = {Phys. Rev. Lett.},
  volume = {127},
  issue = {19},
  pages = {190602},
  numpages = {7},
  year = {2021},
  month = {11},
  publisher = {American Physical Society},
  doi = {10.1103/PhysRevLett.127.190602},
  url = {https://link.aps.org/doi/10.1103/PhysRevLett.127.190602}
}

@article{Mitchison_2019,
   title={Quantum thermal absorption machines: refrigerators, engines and clocks},
   volume={60},
   ISSN={1366-5812},
   url={http://dx.doi.org/10.1080/00107514.2019.1631555},
DOI={10.1080/00107514.2019.1631555},
   number={2},
   journal={Contemporary Physics},
   publisher={Informa UK Limited},
   author={Mitchison, Mark T.},
   year={2019},
   month={4}, pages={164–187} }

@article{brunner_12,
  title = {Virtual qubits, virtual temperatures, and the foundations of thermodynamics},
  author = {Brunner, Nicolas and Linden, Noah and Popescu, Sandu and Skrzypczyk, Paul},
  journal = {Phys. Rev. E},
  volume = {85},
  issue = {5},
  pages = {051117},
  numpages = {14},
  year = {2012},
  month = {5},
  publisher = {American Physical Society},
  doi = {10.1103/PhysRevE.85.051117},
  url = {https://link.aps.org/doi/10.1103/PhysRevE.85.051117}
}

@article{meier2025autonomousquantumprocessingunit,
	author = {Meier, Florian and Huber, Marcus and Erker, Paul and Xuereb, Jake},
	doi = {10.1088/1361-6633/ae7b82},
	journal = {Reports on Progress in Physics},
	month = {jul},
	number = {7},
	pages = {077601},
	publisher = {IOP Publishing},
	title = {Autonomous quantum processing unit: an autonomous thermal computing machine \& its physical limitations},
	url = {https://doi.org/10.1088/1361-6633/ae7b82},
	volume = {89},
	year = {2026}}

@article{Proesmans_2020,
   title={Finite-Time Landauer Principle},
   volume={125},
   ISSN={1079-7114},
   url={http://dx.doi.org/10.1103/PhysRevLett.125.100602},
   DOI={10.1103/physrevlett.125.100602},
   number={10},
   journal={Physical Review Letters},
   publisher={American Physical Society (APS)},
   author={Proesmans, Karel and Ehrich, Jannik and Bechhoefer, John},
   year={2020},
   month=sep }

@article{bennettThermodynamicsComputationReview1982a,
  title = {The Thermodynamics of Computation---a Review},
  author = {Bennett, Charles H.},
  year = 1982,
  month = dec,
  journal = {International Journal of Theoretical Physics},
  volume = {21},
  number = {12},
  pages = {905--940},
  issn = {1572-9575},
  doi = {10.1007/BF02084158}
}

@article{buffoniCooperativeQuantumInformation2023,
  title = {Cooperative Quantum Information Erasure},
  author = {Buffoni, Lorenzo and Campisi, Michele},
  year = 2023,
  month = mar,
  journal = {Quantum},
  volume = {7},
  eprint = {2206.10230},
  primaryclass = {quant-ph},
  pages = {961},
  issn = {2521-327X},
  urldate = {2026-01-14}
}

@article{tarantoExponentialImprovementQuantum2020,
  title = {Exponential Improvement for Quantum Cooling through Finite-Memory Effects},
  author = {Taranto, Philip and Bakhshinezhad, Faraj and Sch{\"u}ttelkopf, Philipp and Clivaz, Fabien and Huber, Marcus},
  year = 2020,
  month = nov,
  journal = {Physical Review Applied},
  volume = {14},
  number = {5},
  eprint = {2004.00323},
  primaryclass = {quant-ph},
  pages = {054005},
  issn = {2331-7019},
  doi = {10.1103/PhysRevApplied.14.054005},
  urldate = {2026-01-14}
}

@article{tarantoEfficientlyCoolingQuantum2025,
  title = {Efficiently {{Cooling Quantum Systems}} with {{Finite Resources}}: {{Insights}} from {{Thermodynamic Geometry}}},
  shorttitle = {Efficiently {{Cooling Quantum Systems}} with {{Finite Resources}}},
  author = {Taranto, Philip and {Lipka-Bartosik}, Patryk and {Rodr{\'i}guez-Briones}, Nayeli A. and {Perarnau-Llobet}, Mart{\'i} and Friis, Nicolai and Huber, Marcus and Bakhshinezhad, Pharnam},
  year = 2025,
  month = feb,
  journal = {Physical Review Letters},
  volume = {134},
  number = {7},
  eprint = {2404.06649},
  primaryclass = {quant-ph},
  pages = {070401},
  issn = {0031-9007, 1079-7114},
  doi = {10.1103/PhysRevLett.134.070401}
}

@article{xuerebCoolingQubitUsing2025,
  title = {Cooling a Qubit Using n Others},
  author = {Xuereb, Jake and Stratton, Benjamin and Rolandi, Alberto and He, Jinming and Huber, Marcus and Bakhshinezhad, Pharnam},
  journal = {PRX Quantum},
  volume = {6},
  issue = {4},
  pages = {040368},
  numpages = {38},
  year = {2025},
  month = {Dec},
  publisher = {American Physical Society},
  url = {https://link.aps.org/doi/10.1103/hrph-dbv7}
}

@article{millerQuantumFluctuationsHinder2020,
  title = {Quantum Fluctuations Hinder Finite-Time Information Erasure near the {{Landauer}} Limit},
  author = {Miller, Harry J. D. and Guarnieri, Giacomo and Mitchison, Mark T. and Goold, John},
  year = 2020,
  month = oct,
  journal = {Physical Review Letters},
  volume = {125},
  number = {16},
  eprint = {2007.01882},
  primaryclass = {quant-ph},
  pages = {160602},
  issn = {0031-9007, 1079-7114},
  doi = {10.1103/PhysRevLett.125.160602},
  urldate = {2026-02-16}
}

@misc{dagoFoolingLandauerBound2026,
  title = {Fooling the {{Landauer}} Bound with a Demon Biased Thermal Bath},
  author = {Dago, Salamb{\^o} and Bellon, Ludovic},
  year = 2026,
  month = feb,
  number = {arXiv:2602.06560},
  eprint = {2602.06560},
  primaryclass = {cond-mat},
  publisher = {arXiv},
  doi = {10.48550/arXiv.2602.06560},
  urldate = {2026-02-10}
}

@misc{ho2023limitsenergyefficiencycmos,
      title={Limits to the Energy Efficiency of CMOS Microprocessors}, 
      author={Anson Ho and Ege Erdil and Tamay Besiroglu},
      year={2023},
      eprint={2312.08595},
      archivePrefix={arXiv},
      primaryClass={cs.ET}
}

@article{Pop2010,
	author = {Pop, Eric},
	date = {2010/03/01},
	doi = {10.1007/s12274-010-1019-z},
	id = {Pop2010},
	isbn = {1998-0000},
	journal = {Nano Research},
	number = {3},
	pages = {147--169},
	title = {Energy dissipation and transport in nanoscale devices},
	url = {https://doi.org/10.1007/s12274-010-1019-z},
	volume = {3},
	year = {2010}}

@article{janovitch2025activequantumreservoirengineering,
  doi = {10.22331/q-2026-06-19-2143},
  url = {https://doi.org/10.22331/q-2026-06-19-2143},
  title = {Active {Q}uantum {R}eservoir {E}ngineering: {U}sing a {Q}ubit to {M}anipulate its {E}nvironment},
  author = {Janovitch, Marcelo and Brunelli, Matteo and P. Potts, Patrick},
  journal = {{Quantum}},
  issn = {2521-327X},
  publisher = {{Verein zur F{\"{o}}rderung des Open Access Publizierens in den Quantenwissenschaften}},
  volume = {10},
  pages = {2143},
  month = jun,
  year = {2026}
}

@article{Xuereb_2023,
   title={Impact of Imperfect Timekeeping on Quantum Control},
   volume={131},
   ISSN={1079-7114},
   url={http://dx.doi.org/10.1103/PhysRevLett.131.160204},
   number={16},
   journal={Physical Review Letters},
   publisher={American Physical Society (APS)},
   author={Xuereb, Jake and Erker, Paul and Meier, Florian and Mitchison, Mark T. and Huber, Marcus},
   year={2023},
   month=Oct }

@BOOK{Ash1990-jz,
  title     = "Information Theory",
  author    = "Ash, R B",
  publisher = "Dover Publications",
  year      =  1990,
  isbn  = 0486665216
}

@misc{gyorfi2012refinementslargedeviationtail,
      title={Some Refinements of Large Deviation Tail Probabilities}, 
      author={Laszlo Gyorfi and Peter Harremoes and Gabor Tusnady},
      year={2012},
      eprint={1205.1005},
      archivePrefix={arXiv},
      primaryClass={math.ST},
      url={https://arxiv.org/abs/1205.1005}, 
}

@article{Evans_2014,
doi = {10.1088/0953-8984/26/10/103202},
url = {https://doi.org/10.1088/0953-8984/26/10/103202},
year = {2014},
month = {feb},
publisher = {IOP Publishing},
volume = {26},
number = {10},
pages = {103202},
author = {Evans, R F L and Fan, W J and Chureemart, P and Ostler, T A and Ellis, M O A and Chantrell, R W},
title = {Atomistic spin model simulations of magnetic nanomaterials},
journal = {Journal of Physics: Condensed Matter}
}

@article{Fiorelli_2023,
   title={Mean-field dynamics of open quantum systems with collective operator-valued rates: validity and application},
   volume={25},
   ISSN={1367-2630},
   url={http://dx.doi.org/10.1088/1367-2630/ace470},
   DOI={10.1088/1367-2630/ace470},
   number={8},
   journal={New Journal of Physics},
   publisher= {IOP Publishing},
   author={Fiorelli, Eliana and Müller, Markus and Lesanovsky, Igor and Carollo, Federico},
   year={2023},
   month=Aug, pages={083010} }

@book{1996_Magnetism_Leccabue_F,
author = {Leccabue, F and Sagredo, V},
title = {Magnetism, Magnetic Materials and Their Applications},
publisher = {WORLD SCIENTIFIC},
year = {1996},
doi = {10.1142/3164},
address = {},
edition   = {},
URL = {https://www.worldscientific.com/doi/abs/10.1142/3164},
eprint = {https://www.worldscientific.com/doi/pdf/10.1142/3164}
}

@book{Taur_Ning_2009, 
place={Cambridge}, 
edition={2}, 
title={Fundamentals of Modern VLSI Devices}, 
publisher={Cambridge University Press}, 
author={Taur, Yuan and Ning, Tak H.},
year={2009}}

@misc{carrascocodina2026energyefficiencyquantumcomputers,
      title={Energy efficiency of quantum computers}, 
      author={Miquel Carrasco-Codina and Pau Escofet and Paul Hilaire and Ariane Soret and Sam Nerenberg and Victor Champain and Gerard Milburn and Klara Theophilo and Sophie H. Li and Irais Bautista and Andrés Gómez and Jose Miralles and Sergi Abadal and Carmen G. Almudéver and Eduard Alarcón and Raja Yehia},
      year={2026},
      eprint={2605.15090},
      archivePrefix={arXiv}
}

@article{Usui_2021,
  title = {Simplifying the design of multilevel thermal machines using virtual qubits},
  author = {Usui, Ayaka and Niedenzu, Wolfgang and Huber, Marcus},
  journal = {Phys. Rev. A},
  volume = {104},
  issue = {4},
  pages = {042224},
  numpages = {14},
  year = {2021},
  month = {Oct},
  publisher = {American Physical Society},
  doi = {10.1103/PhysRevA.104.042224},
  url = {https://link.aps.org/doi/10.1103/PhysRevA.104.042224}
}
\vspace{0.5cm}
\begin{center}
    {\bfseries\large End Matter}
\end{center}

\section{Turning High Dimensional Constituents Into Effective Qubits}
\begin{figure}[h!]
    \centering
    \includegraphics[width=\linewidth, trim={0.0cm 0cm 0.0cm 0cm}]{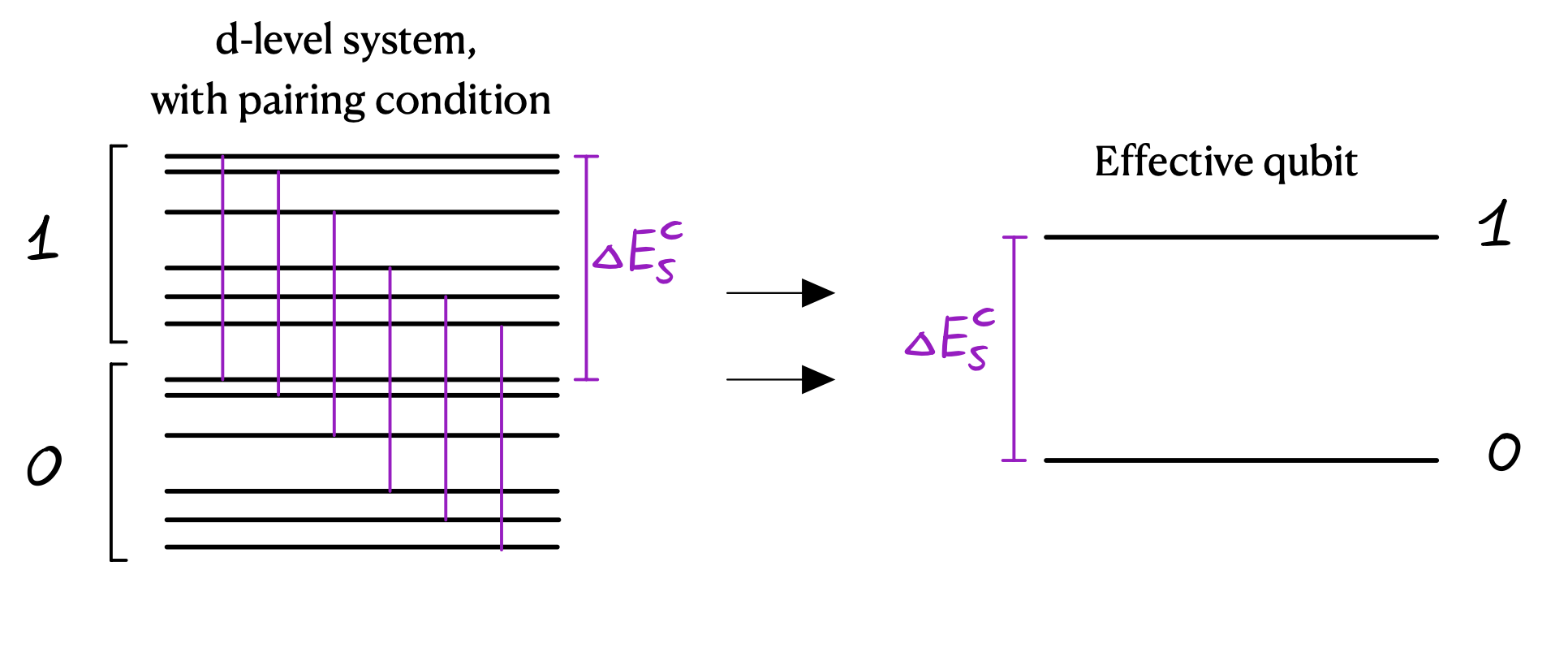}
    \caption{A $d$-level system whose states from the two local subspaces can be paired by a constant gap $\Delta E_{\text{S}}^{\text{C}}$ can be reduced to an effective qubit.}
    \label{fig:EffectiveQubits}
\end{figure}
Here we show explicitly why treating the $N$ constituents of the many-body classical bit as $N$ two-level systems, is not merely a convenient simplification but exact for any constituent whose transitions can be paired by a single control frequency. We do this by proving that the heat dissipated by a single population exchange with the bath and work source for a general $d$-level system to that of a qubit is the same. Consider a single constituent with $d$ levels which we assume may be paired into $d/2$ transitions of equal gap $\Delta E_S^C$, exactly as constructed in the supplemental material~\cite{supmat}. Coupling each pair once to the bath and work source, i.e. performing a single swap per pair ($n=1$ in Eq.~\eqref{eq: p_i'n after n swaps}), starting from the maximally mixed populations $p_i = p_i' = 1/d$, transfers a population $\Delta p = p_{i',1} - 1/d$ into the logical 0 level of each pair. Evaluating Eq.~\eqref{eq: p_i'n after n swaps} at $n=1$ gives
\begin{equation}
    \Delta p = p_{i',1} - \frac{1}{d} = \frac{p_{0,\text{V}}-p_{1,\text{V}}}{d} \, ,
    \label{eq: single pair population transfer}
\end{equation}
independent of which pair we consider, since every pair couples a refreshed virtual qubit. In general for a single population transfer $\Delta p$ the dissipation is given by $\Delta Q_B = \Delta E_B \Delta p$  \cite{supmat} into the bath~\cite{Taranto_2025}, thus by our single pair we dissipate an amount of $\Delta E_B(p_{0,\text{V}}-p_{1,\text{V}})/d$. As all $d/2$ pairs are addressed simultaneously by the same control frequency, the total dissipation of the full $d$-level constituent after a single global swap is just $d/2$ times of that per pair 
\begin{equation}
    \Delta Q_\text{B} = \frac{d}{2} \Delta E_B\left(p_{i',1}-\frac{1}{d}\right) = \Delta E_\text{B} \frac{p_{0,\text{V}}-p_{1,\text{V}}}{2} \, ,
    \label{eq: single swap dissipation d-level}
\end{equation}
where the factor of $d$ cancels exactly between the number of pairs and the population transferred per pair. This result is independent of $d$, thus $d=2$ i.e. a qubit undergoing one swap, the dissipation is the same as for a say $d=100$ constituent. Equivalent erasure of a $d$-level via $d/2$ virtual qubit subspace swaps therefore dissipates exactly as much heat as the erasure of a qubit with the same gap $\Delta E_S^C$, regardless of $d$. This analysis allows us to replace a higher-dimensional system by a single effective qubits. But note that generally, this will not be exactly possible as we need this nice property of being able to pair the states with a constant gap.

\begin{figure}[h!]
    \centering
    \includegraphics[width=\linewidth, trim={0.0cm 0cm 0.0cm 0cm}]{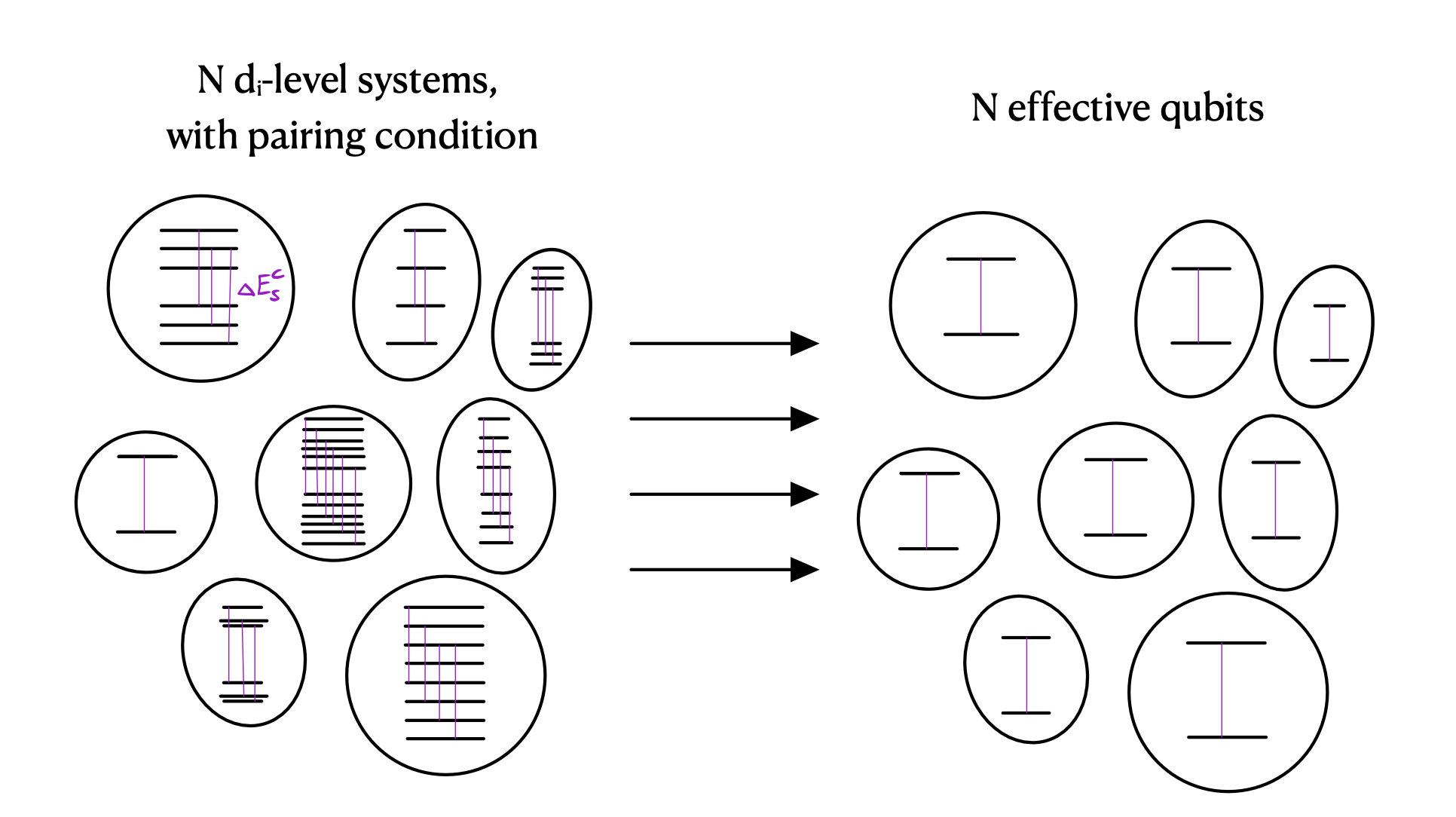}
    \caption{A collection of subsystems of different dimensions, each subject to the pairing condition with the same constant gap $\Delta E_{\text{S}}^{\text{C}}$ can be reduced to a collection of $N$ effective qubits.}
    \label{fig:collectiveQubits}
\end{figure}

Now as we have established that a single $d$-level system is for our task the same as a single qubit, we can reduce all higher-dimensional subsystems of the classical bit to qubits, such that the classical bit can be thought of as a collection of $N$ qubits, see Fig.~\ref{fig:collectiveQubits}. This reduction does not affect the complexity or parallelizability of the protocol, as the $d/2$ pairs of each constituent are addressed simultaneously by the same control frequency, so we can still apply the protocol in parallel for a time $\tau_{\text{opt}}$, see Eq.~\eqref{eq: classical unitary}. Note, however, that the reduction cannot be carried further to the bit as a whole, as the $N$ qubits are not subject to the pairing condition.

Lastly we want to justify the pairing condition and want to outline why it is actually a fair assumption. This is because it just states that we are able to erase the constituent with the given control. If a level of the local 1 subspace has no partner at gap $\Delta E^{\text{C}}_{\text{S}}$, its population is locked and cannot be moved by a single control frequency, so this part of the constituent is inaccessible to erasure altogether. Excluding such constituents is a restriction of our protocol, but a realistic one, since a storage element is only useful if its states can be addressed by the applied control.

\section{Relation to Physical setups}
In this section we want to clarify that the considered setup is general and closely resembles several physical implementations of cooling and erasure. As an illustrative quantum example, consider sideband cooling. Here the system, environment, and laser initially constitute independent resources, while a red-detuned laser provides the frequency-selective work source appearing in Eq.~\eqref{eq: General interaction Hamiltonian}. Increasing the number of accessible bath gaps in our protocol then corresponds to introducing additional laser frequencies or dynamically varying the drive. Although this can improve cooling and subsequently decrease the dissipation, it requires increasingly complicated control.

The finite-time restriction has an equally direct interpretation. A sideband transition occurs at a finite rate set by the effective coupling strength, such that reaching the target state with increasing accuracy requires increasingly long interaction times. Experimental imperfections such as photon losses or rethermalization, which we neglect in our model, would further increase the dissipation, underscoring how idealized our setup is; this is also reflected in Fig.~\ref{fig: combined Plots}(a), where our results lie only one to two orders of magnitude above the Landauer bound. Our protocol therefore captures the basic competition between cooling accuracy, available interaction time, and control complexity, while still neglecting a few imperfections that would push the achievable dissipation further from this bound.

The classical encoding also has a direct physical counterpart. Consider a magnetic memory, where a logical bit is encoded collectively in a magnetic domain and its value is determined by the sign of the total magnetization. The effective two-level subsystems in our model then correspond to microscopic magnetic moments. Importantly, these constituents need not be controlled independently: a global magnetic field can act on the entire ensemble simultaneously, corresponding to the parallel interactions in Eq.~\eqref{eq: classical unitary}. Large subsystem numbers are natural in such encodings; magnetic domains may contain approximately $10^6-10^8$ microscopic spins \cite{Evans_2014}. Thus, increasing $N$ is not merely a mathematical way of improving the protocol, but reflects the intrinsic redundancy of macroscopic classical information storage. The same idea applies to other collective encodings, such as a bit defined by the charge stored on a floating gate.

Finally, the repeated swaps used throughout our analysis should not be interpreted as requiring the bath and work source to be physically replaced after every interaction. They provide a microscopic description of thermalization with a large reservoir. Under continuous coupling, the same interaction leads to relaxation towards the virtual temperature, with the virtual-qubit populations determining the effective thermalization rate \cite{brunner_12, Usui_2021}. Consequently, the finite number of swaps considered here can equivalently be viewed as a finite available thermalization time.

\newpage
\pagebreak

\appendix
\onecolumngrid

\newpage
\begin{center}
\vskip0.5cm
{\Large Supplemental Material:}
\vskip0.2cm
{\large Quantum vs Classical Erasure: Equal Bounds but Unequal Costs}
\end{center}
\setcounter{equation}{0}
\setcounter{figure}{0}
\setcounter{table}{0}
\setcounter{page}{1}
\setcounter{section}{0}
\renewcommand{\thesection}{S\arabic{section}}
\renewcommand{\theequation}{S\arabic{equation}}
\renewcommand{\thefigure}{S\arabic{figure}}

\section{Proof of the Equivalent Landauer Bound of Classical and Quantum Systems}
For a process involving a system and a thermal reservoir going from initially uncorrelated states $\rho_{\text S/R}$ to $\rho'_{\text SR}$ it holds that dissipation into the bath is given by \cite{Reeb_and_Wolf}
\begin{equation}
     \beta \Delta Q_{\text B} = \mathcal{I}(S';R') + D(\rho'_{\text R}||\rho_{\text R}) - \Delta S \, ,
     \label{eq: reeb wolf}
\end{equation}
where the mutual information $\mathcal{I}(S';R') = S(\rho_{\text S}') + S(\rho_{\text R}') - S(\rho_{\text SR}')$ quantifies the correlation between the system and environment after the process, with the von Neumann entropy given by $S(\rho) = - \text{tr}\{ \rho \log(\rho)\}$ and $\rho_{\text S/R}' = \tr_{R/S} \{\rho_{\text SR}'\}$. Further the relative entropy $D(\rho'_{\text R}||\rho_{\text R}) = \tr\{\rho'_{\text R} \log(\rho'_{\text R})\}-\tr\{\rho'_{\text R} \log(\rho_{\text R})\}$ measures how much the reservoir state was perturbed. Notably both the mutual information and the relative entropy are  positive quantities, therefore we find the Clausius inequality $\beta \Delta Q_{\text B} \ge - \Delta S$, where  $\Delta Q_{\text B} = \text{tr}\{H_{\text R}(\rho'_{\text R} - \rho_{\text R})\}$ is the heat change or dissipation of the thermal reservoir and $\Delta S = S(\rho_{\text S}') - S(\rho_{\text S})$  is the entropy change of the system. Note that $\beta \Delta Q_{\text B} \ge - \Delta S$ also holds for energy-conserving unitaries and if we use two thermal baths to drive the cooling it gets prefaced with the Carnot factor \cite{Taranto_2023}.

Cooling a quantum bit to a ground state population of $\mathcal{F}$ and a population of $1-\mathcal{F}$ in the excited state we find the change in entropy, which gives the minimal dissipation for finite erasure of a qubit by 
\begin{equation}
    \beta \Delta Q_{\text B}^{\text (Q)} \geq  \log(2)+\mathcal{F} \log (\mathcal{F}) +(1-\mathcal{F})  \log (1-\mathcal{F}) \, .
    \label{eq: finite landauer bound}
\end{equation}

Next we consider the finite erasure in a $d$-dimensional system, where we cool it to a state with a distributed probability of $\mathcal{F}$ in the $d/2$ states corresponding to the logical 0 and $1-\mathcal{F}$ in the other $d/2$, thus we may write $\rho'_{d}  = \sum^{d/2}_{j = 1} q_j \ketbra{j}{j} + \sum^{d}_{j = d/2+1} q'_j \ketbra{j}{j}$, where the $q_j,q_j'\geq 0$ are probabilities that have to sum to $\mathcal{F}$ and $1-\mathcal{F}$ and the $\ket{j}$ are some eigenstates of an arbitrary operator, with this we find 
\begin{equation}
    \beta  \Delta Q_{\text B}^{\text (C)} \ge \log(d) + \sum_{j = 1 }^{d/2} q_j \log(q_j) + \sum_{j = d/2+1 }^{d} q'_j \log(q'_j) \, .
\end{equation}
Which at best coincides with the quantum lower bound, if we distribute the probability uniformly in each subspace i.e. $q_j' = (1-\mathcal{F}) \times 2/d$ and $q_j = \mathcal{F}\times 2/d$, for all $j$. Note that for perfect erasure ($\mathcal{F} = 1$) we retrieve the famous Landauer bound of $\beta\Delta Q \geq \log(2)$ in both cases.

We emphasize that these results are independent of the detailed structure and energy spectrum of the $d$-dimensional (or qubit) system and depend only on the amount of information erased or redistributed. Consequently, this limit provides limited insight into the underlying physical mechanisms of the erasure process.

\section{Derivation of the Population Changes of the Protocol}
To formalise the interaction with the thermal bath and work source, we adopt the concept of a \emph{virtual qubit} as in for example Refs.~\cite{brunner_12,Clivaz_2019}. For this we identify a two-dimensional subspace of the joint bath and work source Hilbert space, defined by $\ket{0_{\text V}} = \ket{0_{\text B} 1_{\text W}}$ and $\ket{1_{\text V}} = \ket{1_{\text B} 0_{\text W}}$, chosen such that they satisfy the resonance condition of Eq.~\eqref{eq: energy resonance condition}. This subspace behaves as an effective two-level system with energy gap $\Delta E_{\text S}^{\text Q,C}$, whose population ratio defines a virtual temperature $\beta_{\text V}$ \cite{brunner_12}. Importantly, this virtual qubit is generally not normalised, i.e. the populations of it are less than one, reflecting that these are populations of a larger Hilbert space. Further coherent interactions that swap population between the system and this subspace are equivalent to thermalization to this virtual temperature.

Now we want to use this virtual qubit picture in our setup. First of interacting with the thermal bath plus work source is the same as swapping coherently with the virtual qubit. Applying the tripartite interaction given by the Hamiltonian Eq.~\eqref{eq: General interaction Hamiltonian} by the unitary $U(\tau) = e^{-iH\tau}$ we can find how the populations transform in time by 
\begin{equation}
  p_{i'} \rightarrow  \tr \left\{\ketbra{E_{i'}^\text{S}}{E_{i'}^\text{S}}U(\tau) \rho_{\text Q, C} \otimes \tau_{\beta} \otimes \rho_{\text W} U^\dagger(\tau) \right\} \, , 
\end{equation}
which leads you to the transformation of 
\begin{equation}
p_{i'} \rightarrow p_{i'} + \sin^2(g \tau )\frac{ \left(p_i w_1 e^{-\beta  E^{\text S}_{i'}}-p_{i'} w_0 e^{-\beta E^{\text S}_i}\right)}{\mathcal{Z}_\text{B}} \, , 
\end{equation}
where $\mathcal{Z}_\text{B} = \tr \{e^{-\beta H_{\text B}}\}$ is the partition function of the bath, further as $p_{i} +p_{i'} = 1 $ we can find $p_{i}$ straightforwardly by $1-p_{i'}$. For the optimal time (maximal population $p_{i'}$) of $\tau_{\text opt} = \pi/(2 g)$ this will transfers populations of the systems as
\begin{align}
    p_i \rightarrow p_i - \Delta p && 
    p_{i'} \rightarrow p_{i'} +\Delta p \, , 
\end{align}
where $p_i$ is the population of energy state $\ket{E^\text{S}_i}$, further $\Delta p$ is given by
\begin{equation}
    \Delta p = p_{i} p_{0, \text{V}}-p_{i'} p_{1,\text{V}}  \, , 
\end{equation}
where we defined the virtual qubit populations
\begin{align}
      p_{0,\text{V}} =  \langle 0_{\text B} 1_{\text W}|\rho_{BW}|0_{\text B}1_{\text W}\rangle  = \frac{ w_1 e^{-\beta  E^{\text S}_{i'}}}{\mathcal{Z}_\text{B}}
      &&
    p_{1,\text{V}} =\langle 1_{\text B}0_{\text W}|\rho_{BW}|1_{\text B}0_{\text W}\rangle= \frac{ w_0 e^{-\beta  E^{\text S}_{i}}}{\mathcal{Z}_\text{B}}
    \, .
\end{align}
which are the populations of the energy eigenstates corresponding to $H_{B/W}|0/1\rangle=E^{B/W}_{i/i'}|0/1\rangle$.

The populations of the target system after $n$ such interactions may then be found via a recursive formula 
\begin{equation}
    p_{i'}^{(n)} = (p_i + p_{i'}) \frac{1-(1-p_{0,\text{V}}-p_{1,\text{V}})^n}{1 +\frac{p_{1,\text{V}}}{p_{0,\text{V}}}} + p_i (1-p_{0,\text{V}}-p_{1,\text{V}})^n \, , 
    \label{eq: p_i'n after n swaps}
\end{equation}
where $ p_{i'}^{(n)}$ is the population of the $i$ energy level of the system after $n$ perfectly timed swaps. By now assuming that the virtual qubit subspace is formed of a thermal reservoir qubit at temperature $\beta$ with energetic gap $\Delta E_{\text B}$ and a work source with a pair of energy levels with populations $w_0, w_1 \, : \, w_0> w_1$, we obtain the explicit form 
\begin{equation}
    p_{i'}^{(n)} = \frac{(p_i+p_{i'}) \left(1-\left(1-w_1\frac{1+\frac{w_0}{w_1} e^{-\beta  \Delta E_{\text B}}}{\mathcal{Z}_{\text B}}\right)^n\right)}{1+ \frac{w_0}{w_1} e^{-\beta  \Delta E_{\text B} }}
    +p_{i'} \left(1-w_1\frac{1+\frac{w_0}{w_1} e^{-\beta  \Delta E_{\text B} }}{\mathcal{Z}_{\text B}}\right)^n \, .
    \label{eq: Populiation of pi after n swaps}
\end{equation}
As the virtual qubit is not normalized $\abs{1-p_{0,\text{V}}-p_{1,\text{V}}} <1 $, finding the population in the limit of asymptotically many population exchanges is simply
\begin{equation}
    \lim_{n \rightarrow \infty } p_{i'}^{(n)} = \frac{p_i+ p_{i'}}{1 +\frac{p_{1,\text{V}}}{p_{0,\text{V}}}} = \frac{p_i+ p_{i'}}{1+e^{-\beta\Delta E_{\text B}} \mathcal{W} } \, . 
\end{equation}
This coincides with the target two-level subspace having thermalized with the virtual qubit temperature \cite{brunner_12}
\begin{equation}
 \beta_{\text V}  = \frac{\beta\Delta E_{\text B} - \log( \mathcal{W}) }{\Delta E _{\text S} ^{\text Q,C}} \, , 
 \label{eq: virtual bath temp}
\end{equation}
and gap $\Delta E _{\text S} ^{\text Q,C}$, which is needed in order to make the interaction energy preserving as we stress in the main body is needed for a full thermodynamic accounting. By Eq.~\eqref{eq: virtual bath temp} we are able to cool to a colder state, by either interacting with a colder bath given by $\beta$ or a larger energetic gap $\Delta E_{\text B}$ whilst the work source merely constrains the achievable populations for fixed $\beta, \Delta E_{\text B}$.

\section{How to Calculate the Dissipation} 

In this section we show how to find the dissipation $\Delta Q_\text{B}$ for a single unitary interaction $U$ with the virtual qubit subspace, which is defined through $\Delta Q_{\text B,j} = \text{tr}\{H_{\text B}(U\rho_{\text tot}U^\dagger-\rho_{\text tot})\}$, where $U$ is generated by a single $H_j$ given by Eq.~\eqref{eq: General interaction Hamiltonian}. Since this interaction only changes populations in the energy eigenbasis, or equivalently the levels we have picked to cool, we find 
\begin{equation} \label{eq: Dissipation formula single swap} 
    \Delta Q_{\text B, j } = E^{\text B}_i \Delta p_{\text B, i} + E^{\text B}_{i'} \Delta p_{\text B, i'} = \Delta E_B \Delta p_{\text{S},i}^{(j)} \, ,
\end{equation}
where $\Delta p_{\text B, i}= \text{tr}\{\ketbra{E^{\text B}_i}{E^{\text B}_i}(U\rho_{\text tot}U^\dagger-\rho_{\text tot})\}$ and $\Delta p_{\text{S}, i}^{(j)}= \text{tr}\{\ketbra{E^{\text S}_i}{E^{\text S}_i}(U\rho_{\text tot}U^\dagger-\rho_{\text tot})\}$ are the population changes of the local logical 0 state we have cooled~\cite{Taranto_2025}, and we have added the label $j$ to $\Delta p_{\text{S},i}^{(j)}$ to track which swap/virtual qubit it belongs to.

We now extend this to a protocol of several swaps, each swap in the sequence couples the system to a freshly refreshed bath-work virtual qubit, but always across the same resonance condition, Eq.~\eqref{eq: energy resonance condition}, so every swap dissipates at the same fixed gap $\Delta E_\text{B}$, thus only the amount of population transferred, $\Delta p_{\text{S},i}^{(j)}$, can differ from swap to swap. Since Eq.~\eqref{eq: Dissipation formula single swap} gives the dissipation of an individual swap $j$, the total dissipation after $n$ swaps is simply their sum,
\begin{equation}
    \sum_{j=1}^n \Delta Q_{\text B, j} = \Delta E_\text{B} \sum_{j=1}^n \Delta p_{\text{S},i}^{(j)} \, .
\end{equation}
The key simplification is that $\Delta p_{\text{S},i}^{(j)} = p_i^{(j)} - p_i^{(j-1)}$ is precisely the change in the system's population from immediately before swap $j$ to immediately after it, where $p_i^{(0)} = p_i$ is the initial population and $p_i^{(n)}$ is the population after all $n$ swaps. The sum over $j$ is therefore telescoping and can be simplified to
\begin{equation}
    \sum_{j=1}^n \Delta p_{\text{S},i}^{(j)} = \sum_{j=1}^n \left(p_i^{(j)} - p_i^{(j-1)}\right) = p_i^{(n)} - p_i^{(0)} = \Delta p_{\text{S}, i }\, .
\end{equation}
Consequently, however many swaps the protocol uses, the total dissipation depends only on the net population moved from the initial to the final state and the gap $\Delta E_B$ used
\begin{equation}
    \Delta Q_\text{B} = \sum_{j=1}^n \Delta Q_{\text{B},j} = \Delta E_\text{B} \, \Delta p_{\text{S}, i } \, .
\end{equation}

This might suggest that the total dissipation depends only on this net population change, independent of the specific protocol used to reach it. This is not quite the case, for a fixed protocol, the achievable $\Delta p_{\text{S}, i }$ is itself set by the gap $\Delta E_\text{B}$, as captured for example by Eq.~\eqref{eq: infintie swap limit p_i} in the asymptotic-swap case. We can make this connection explicit using Eq.~\eqref{eq: rescaled dissipation by fidelity}. Solving Eq.~\eqref{eq: infintie swap limit p_i} for the gap gives \begin{equation} \Delta E_\text{B} = \frac{1}{\beta} \log\left( \frac{F}{1-F}\mathcal{W}\right) \, , \end{equation} while expressing the population change in terms of the fidelity gives $\Delta p_{\text{S}, i }= F(p_i+p_{i'}) - p_i = (F-1)p_i + Fp_{i'}$. Substituting both into $\Delta Q_\text{B} = \Delta E_\text{B}\Delta p_{\text{S}, i }$ we find Eq.~\eqref{eq: rescaled dissipation by fidelity}. 

Similarly, one finds Eq.~\eqref{eq: Divpoint qubit} by taking the limit $\Delta E_\text{B} \rightarrow \infty$ of Eq.~\eqref{eq: Populiation of pi after n swaps}. This limit corresponds to the largest erasure achievable with $n$ swaps, but only at the cost of infinite dissipation, as $\Delta p_{\text{S},i}$ approaches a finite value while $\Delta E_\text{B}$ diverges, $\Delta Q_\text{B} = \Delta E_\text{B}\Delta p_{\text{S},i}$ diverges as well. At the opposite extreme $\Delta E_\text{B} = 0$ induces no population transfer at all i.e.\ $\Delta p_{\text{S},i} = 0$ and hence $\Delta Q_\text{B} = 0$, leaving the system at its initial fidelity $\mathcal{F} = 1/2$. Note that $\Delta Q_\text{B}$ depends continuously on $\Delta E_\text{B}$, thus it follows that for any fixed number of swaps $n$, $\Delta Q_\text{B}$ is a continuous function of the fidelity $\mathcal{F}$ between $\Delta Q_\text{B} = 0$ at $\mathcal{F} = 1/2$ and $\Delta Q_\text{B} \to \infty$ as $\mathcal{F} \to \mathcal{F}^{\text{max}}_\text{Q}(n)$. This reveals the trade-off the qubit faces between swap number and dissipation, to reach a given fidelity $\mathcal{F}$, using more swaps $n$ allows the same erasure to be achieved with a smaller gap $\Delta E_\text{B}$, and hence at a lower dissipation $\Delta Q_\text{B}$.

\section{Similar Dissipation of $d$-level Systems with Pairing Condition and Qubit systems}
In this section we provide a constructive protocol for the erasure of a bit of information encoded in a $d$-level system, under the assumption that its energy levels can be paired as required by our single control frequency and we further show that this does not come at a larger cost than in the qubit case. Assume our system has a Hamiltonian 
\begin{equation}
    H = \sum_{n =1}^d E_n\ketbra{n}{n} \, , 
\end{equation}
where we assume that the $E_n$ are sorted increasingly i.e. $E_1 \leq E_2\leq \cdots \leq E_d$ and $d$ is even, so that the logical 0 and 1 subspaces contain the same number of microstates. We label the lower half of the energy states as the 0 subspace i.e. $\Pi_0 = \sum_{n = 1}^{d/2} \ketbra{n}{n}$, which is in agreement with our assumption that half of all the states are in each logical space. To erase we pick pairs of each subspace, which we want to move populations in, we pick now $\ket{n}\leftrightarrow\ket{\frac{d}{2}+n} $, where we assume that these energy states have a constant energy gap of $\Delta E_{\text S}^C $, thus a single control frequency of the bath and work source is enough to allow for erasure in this system. Note that this condition can be realised by several level structures, examples include an equally spaced ladder, for which $\Delta E^C_S = \frac{d}{2} \omega$, where $\omega$ is the spacing of the ladder or a $d/2$-fold degenerate two-level structure, where all pairs trivially share one gap. More generally, any $d$-level system whose spectrum happens to admit such a pairing, degenerate or not can be erased in this manner.

Now we employ Eq.~\eqref{eq: rescaled dissipation by fidelity} on these $d/2$ pairs, as we want to start maximally mixed we know $p_i = p_i' = 1/d$ and as we do $d/2$ swaps we find that the total dissipation becomes 
\begin{equation}
    \Delta Q^{\text (C)}_{\text B} = \frac{d}{2} \times \frac{2F-1}{\beta d} \log ( \frac{F}{1-F}\mathcal{W}) \, , 
    \label{eq: Delta Q_B^C}
\end{equation}
where the swap fidelity is defined as
\begin{equation}
    F = \frac{p_i^*}{\frac{1}{d}+ \frac{1}{d}} \Leftrightarrow p_i^*  = \frac{2}{d}F \, , 
\end{equation}
and $p_i^*$ denotes the population of the 0 subspace level belonging to that pair after the swap. Since every pair is swapped identically, $p_i^*$ takes the same value for each of the $d/2$ pairs, so the total population of the logical 0 subspace follows directly by summing over all pairs
\begin{equation}
      \mathcal{F}  =\tr\{\Pi_0\rho'\} = \sum_{n = 1}^{d/2} p_i^* = \frac{d}{2}\frac{2}{d}F \Leftrightarrow F = \mathcal{F} \,.
\end{equation}
Thus the erasure fidelity $\mathcal{F}$ is the same as the swap fidelity $F$, as in the qubit case, precisely the total logical 0 subspace population. The dissipation incurred in erasing this $d$-level system, Eq.~\eqref{eq: Delta Q_B^C}, therefore coincides exactly with the dissipative cost of erasing the qubit, see Eq.~\eqref{eq: rescaled dissipation by fidelity} for $p_i = p_i' = 1/2$.

\section{Classical Bits can Counteract insufficient Cooling by an Increased Subsystem Number}
Here we provide the proof that in the partial thermalization setting a classical system is able to counteract the limited population change by a larger subsystem number $N$. For this we firstly have to find how to calculate the fidelity in this setting, for this we work in the setting of $N$ qubit constituents, therefore we can define the logical 0 space projector as 
\begin{align}
    \Pi_0 &= \sum_{s \in \mathcal{S}} \ketbra{s}{s} \, ,& 
    \mathcal{S} &= \left\{ v \in \{0,1\}^N \; \middle| \; \sum_{i=1}^N v_i < \frac{N}{2} \right\} \, .
\end{align}
Since the $N$ constituents are cooled independently and identically, the erased state factorises as $\rho' = \bigotimes_{i=1}^N \rho_i'$, with each $\rho_i'$ having the same population $p$ in its local 0 level. The probability of obtaining a given bitstring $v \in \{0,1\}^N$ under $\rho'$ then depends only on how many of its entries are 0 or 1, namely it is $p^{N-\sum_i v_i}(1-p)^{\sum_i v_i}$. Evaluating $\text{tr}\{\Pi_0 \rho'\}$ therefore no longer requires summing over the individual bitstrings in $\mathcal{S}$, as every bitstring with the same number of 0s carries the same probability, we only need to know how many bitstrings of length $N$ have exactly $n$ zeros and multiply this times its probability. This is precisely the binomial coefficient $\binom{N}{n}$, so the number of 0s across the $N$ constituents follows a binomial distribution. Thus $\text{tr}\{\Pi_0\rho'\}$ is the probability that fewer than $N/2$ constituents are found in their local 1 state, giving

\begin{equation}
    \mathcal{F}_{\text C}(N) = \text{tr}\{\Pi_0 \rho'\} = 1- \sum_{n = 0}^{\frac{N-1}{2}} \binom{N}{n}(1-p)^{N-n}p^n  \, .
    \label{eq: Fidelity classical by binom distr}
\end{equation}
Then Eq.~\eqref{eq: Fidelity classical by binom distr} is the probability that at half or more of the effective two level systems are contained in the logical 0-subspace. The binomial distribution of Eq.~\eqref{eq: Fidelity classical by binom distr} can be upper bound by \cite{Ash1990-jz}
\begin{equation}
\mathcal{F}_{\text C}(N)\geq1- e^{-N D(1/2||p)} \xrightarrow{N\rightarrow \infty}1 \, ,
\label{eq: Exponential decrease in error of N}
\end{equation} 
where $\mathcal{D}(a||b) = a \log(\frac{a}{b})+ (1-a)\log(\frac{1-a}{1-b})$ is the relative entropy of two Bernoulli distributions, note that the limit of Eq.~\eqref{eq: Exponential decrease in error of N} is true for any  $p>1/2$. Note that the condition of $p>1/2$ is achieved after interacting for an arbitrary time with a virtual qubit with an arbitrarily small thermal gap $\Delta E_{\text B}$. Thus we do not need to cool each qubit in the classical system to a super pure state, but can actually just cool each qubit minimally and then a single interaction with the bath will be enough, as long as the system is composed of enough subsystems this behavior is able to counteract our cooling limitations. 

Again comparing to a hard drive disk one usually has $10^6- 10^8$ spins per magnetic domain \cite{Evans_2014}, so a large subsystem number is a fairly realistic assumption.

\section{Equal Divergence Point: $N_{\text{ min}}$ Derivation}
Here we provide a quick outline of how we found the approximation of $N_{\text min}$, firstly we know that the reachable population of each subsystem of the 0-subspace after a single interaction is given by $p_{0,i} = \frac{1+w_1}{2}$, stemming from Eq.~\eqref{eq: Divpoint qubit} for $n = 1$, which one obtains from Eq.~\eqref{eq: Populiation of pi after n swaps} by taking the limit $\Delta E_{\text B} \rightarrow \infty$. We will assume an odd $N$ thus we find that the divergence point of the classical system with $N$-subsystems is given by 
\begin{equation}
    \mathcal{F}_{\text C}^{\text max}(N) = \sum_{m = 0}^{\frac{N-1}{2}} \binom{N}{m}\left(\frac{1+ w_1}{2}\right)^{N-m}\left(\frac{1-w_1}{2}\right)^m  \, , 
\end{equation}
which is known to be well approximated for large $N$ by the Stirling's formula \cite{Ash1990-jz} resulting in 
\begin{equation}
    \mathcal{F}_{\text C}^{\text max}(N) \approx1- \frac{1}{\sqrt{8\frac{N-1}{2} (1-\frac{N-1}{2N})}} e^{-N \mathcal{D}\left(\frac{N-1}{2N} || \frac{1+w_1}{2} \right)} =1-\frac{\left(\frac{N-1}{N (w_1+1)}\right)^{\frac{1-N}{2}} \left(\frac{N+1}{N(1- w_1)}\right)^{-\frac{N+1}{2}}}{\sqrt{2\frac{N^2-1}{N}}} \, , 
\end{equation}
where the relative error of this approximation is $\mathcal{O}\left( \frac{1}{N}\right)$ \cite{gyorfi2012refinementslargedeviationtail}. Next we would need to solve $\mathcal{F}_{\text{C}}^{\text max}(N_{\text min}) = \mathcal{F}_{\text{Q}}^{\text max}(n)$ for $N$, but this is not possible as we are dealing with a transcendental equation, therefore we opt to more approximations (assuming large $N$), which is 
\begin{equation}
    \mathcal{F}_{\text C}^{\text max}(N) \approx 1-\frac{\left(\frac{1}{1-w_1}\right)^{-\frac{N}{2}} \left(\frac{1}{1+w_1}\right)^{-\frac{N}{2}}}{\sqrt{2N}} \, . 
\end{equation}
This is enough to solve the equality and we obtain  
\begin{equation}
    N_{\text min} = -\frac{W\left(-2 (1-w)^{-2 n} (\log (1-w)+\log (w+1))\right)}{\log (1-w)+\log (w+1)} \, , 
\end{equation}
where $W$ is the lambert $W$ function, but note that this is under the assumption of large $N$, thus we can also safely assume $n$ to be fairly large, therefore in the limit of $1/n \to 0$ we find 
\begin{equation}\label{eq: minimum N for same divergence point}
    N_{\text min} = \frac{2 n \log (1-w_1)}{\log \left(1-w_1^2\right)} + \frac{\log (n)}{\log \left(1-w_1^2\right)} + \frac{\log \left(\frac{\log (1-w_1)}{\log \left(1-w_1^2\right)}\right)}{\log \left(1-w_1^2\right)}+\mathcal{O}\left(\frac{1}{n}\right) \, .
\end{equation}
This nicely displays that, at leading order, $N_{\text{min}}$ scales linearly in the number of swaps $n$, highlighting the trade-off between the swaps afforded to the quantum system and the subsystem number $N$ of the classical bit. That $N$ and $n$ are large is well justified physically, as detailed in the end matter. For example sideband cooling requires a large photon flux to realise many swaps, while typical classical storage devices, are built from a correspondingly large number of subsystems.

\section{Effects of Imperfect Timekeeping or Missing Knowledge of the Coupling Constant}
In this section we assume that we do not have access to perfect knowledge of the coupling constant $g$ or equally we could have an imperfect clock. We model this by timing each interaction by $\tau = \frac{\pi}{2g} + \frac{\varepsilon}{g}$,  where we draw $\varepsilon$ from a normal distribution $\varepsilon = \mathcal{N}(0,s^2)$, with mean $0$ and variance $s^2$ with $s\ll 1$. We  thus find the local 0 subspace population per subsystem $i$, after a single swap to be 
\begin{equation}
p_{i,0}(\varepsilon) = \frac{1}{2} + w_1\frac{1-\frac{w_0}{w_1} e^{-\beta \Delta E_{\text B}}}{2 \mathcal{Z}_{\text B}}\left(1-\varepsilon^2\right) + \mathcal{O}\left(\varepsilon ^3\right)  \approx  p_{i,0} - c  s^2\chi_1^2 \, , 
\label{eq: 0 subspace population mistiming}
\end{equation}
where $p_{i,0}$ is the optimal 0 subspace population and $ c = w_1\frac{1-\frac{w_0}{w_1} e^{-\beta \Delta E_{\text B}}}{2 \mathcal{Z}_{\text B}}$ and finally $\chi^2_1$ is the chi squared distribution of a normal distributed variable $Z = \mathcal{N}(0,1) $. 

Based on this we want to know the effect of this for the classical erasure protocol we can now either assume that we draw once from the distribution $\varepsilon$, but then all $N$ subsystems of the classical bit will be perturbed to the exact level of erasure, but this gives no new insights as then based on the earlier section this does not change the exponential decrease of the error in $N$ (see Eq.~\eqref{eq: Exponential decrease in error of N}). Therefore we want to focus on the case where for each subsystem $i$ we draw a new sample $\varepsilon_i$ and each local 0 subspace population is then $p_{i,0}(\varepsilon_i)$, by Eq.~\eqref{eq: 0 subspace population mistiming}. This would coincide to the case of applying the protocol for a finite time but the coupling constants to each subsystem are randomly distributed. An example would include not all spins coupling in the same manner to a magnetic field, because of some sort of inhomogeneity \cite{janovitch2025activequantumreservoirengineering}.

Because we now draw $N$ samples the majority voting is a bit more complicated to obtain, but we can use the central limit theorem in order to approximate the fidelity and its scaling. For this we define 
\begin{align}
    S &=  \sum_i X_i \, ,\text{with}  &     X_i &= \begin{cases}
        1 \;\text{with probability }\;  1-p_{i,0}(\varepsilon_i)\\
        0 \;\text{with probability }\;  p_{i,0}(\varepsilon_i)\\
    \end{cases} \, , 
\end{align}
now we employ the central limit theorem on this to find 
\begin{equation}
    S \approx \mathcal{N}(N\mathbb{E}(X_i) , N\text{Var}(X_i)) \, , 
\end{equation}
where 
\begin{equation}
    \mathbb{E}(X_i) =1- \left( p_{i,0} - c  s^2\right)  \, , 
\end{equation}
and further 
\begin{equation}
    \text{Var}(X_i)  = ( 1-p_{i,0})(p_{i,0}- 2c s^2) + c s^2- 5(c s^2)^2 \, , 
\end{equation}
thus we can define the fidelity by 
\begin{equation}
    \mathbb{E}(\mathcal{F}_{\text C}(N,\varepsilon)) = P\left(S\leq\frac{N}{2}\right) = \Phi(N/2, N\mathbb{E}(X_i) , N\text{Var}(X_i)) = \frac{1}{2}\left[1+ \text{erf}\left(\frac{\sqrt{N}}{2}\frac{1- 2\mathbb{E}(X_i) }{\sqrt{2\text{Var}(X_i)} }\right)\right]\, , 
\end{equation}
where $\Phi$ is the cumulative error function and $\text{erf}$ is the Gauss error function, which for large $x$ can be approximated by
\begin{equation}
\text{erf}(x) =  1-\frac{e^{-x^2}}{x} \left(1+ \mathcal{O}\left( \frac{1}{x^2}\right)     \right) \, , 
\end{equation}
implying that still we have a good large $N$ approximation of the fidelity by
\begin{equation}
    \mathbb{E}(\mathcal{F}_{\text C}(N,\varepsilon)) \simeq 1- \frac{e^{-\frac{N}{4}  \frac{(1- 2\mathbb{E}(X_i))^2 }{2\text{Var}(X_i)}}}{\sqrt{N}\frac{1- 2\mathbb{E}(X_i) }{\sqrt{2\text{Var}(X_i)} }} \, , 
    \label{eq: Exponential decrease classical for mistimings}
\end{equation}
which still displays the exponential decay in the total subsystem number $N$, even though we allow for mistiming or some missing knowledge of the coupling constant $g$.

Now lets do the same with the qubit for this we see that after $n$ mistimed swaps we obtain 
\begin{equation}
    \mathcal{F}_{Q}(n, \varepsilon) =   p_0 \prod_{i = 1}^n c(\varepsilon_i) + \sum_{i = 1}^n k(\varepsilon_i) \prod_{j = i+1}^n c(\varepsilon_j) \, , 
\end{equation}
where 
\begin{align}
    k(\epsilon) &= \frac{w_1}{\mathcal{Z}_{\text B}}\left( 1-\varepsilon^2\right)\, , &
    c(\varepsilon) &=1- w_1 \frac{1+ \frac{w_0}{w_1} e^{-\beta \Delta E_{\text B}}}{\mathcal{Z}_{\text B}} (1-\varepsilon^2)\, , 
\end{align}
and because we do iid sampling we can make use of the fact that the expectation values are independent thus
\begin{equation}
    \mathbb{E}(\mathcal{F}_{\text Q}(n,\epsilon)) =  \frac{\mathbb{E}(c(\varepsilon))^n}{2} + \mathbb{E}(k(\varepsilon))  \sum_{i = 1}^n\mathbb{E}(c(\varepsilon) )^{n-i} = \frac{\mathbb{E}(c(\varepsilon))^n}{2}  + \mathbb{E} (k(\varepsilon)) \frac{1- \mathbb{E}(c(\varepsilon) )^n}{1-\mathbb{E}(c(\varepsilon) )} \, , 
\end{equation}
where we used that the second part is a geometric series. Further
\begin{align}
   \mathbb{E}( k(\epsilon) )&= \frac{w_1}{\mathcal{Z}_{\text B}}\left( 1-s^2\right)\, , &
    \mathbb{E}(c(\epsilon) )&=1- w_1 \frac{1+ \frac{w_0}{w_1} e^{-\beta \Delta E_{\text B}}}{\mathcal{Z}_{\text B}} (1-s^2)\, , 
\end{align}
Now taking the limit of $\Delta E_{\text B} \rightarrow \infty$ and consequently $\mathcal{Z}_{\text B} \rightarrow1$ (either all non ground states also get infinite gaps or we assume a thermal qubit) we obtain that the maximal reachable fidelity of the qubit becomes 
\begin{equation}
    \mathbb{E}(\mathcal{F}_{\text Q}(n,\epsilon)) \xrightarrow{\Delta E_{\text B}\to \infty} 1- \frac{\left(1-(1-s^2)w_1 \right)^n}{2} \, , 
    \label{eq: reachable p0 qubit mistimings}
\end{equation}
therefore in the finite swap regime the total reachable fidelity is limited by the mistiming, which was to be expected, but note that in the limit of infinite swaps we retrieve
\begin{equation}
    \lim_{n \rightarrow \infty} \mathbb{E}(\mathcal{F}_{\text Q}(n,\epsilon)) = \frac{1}{1+ \frac{w_0 }{w_1} e^{-\beta \Delta E_{\text B}}} \, , 
\end{equation}
so in the infinite time case the mistimings do \emph{not} matter recovering the insight of~\cite{Baumer2019imperfect,Xuereb_2023}. As per swap one is able to move less population than in the not mistimed case, but still there is some population moved, in the case of infinite swaps this still will bring you to thermalization to the virtual qubit bath temperature.

\end{document}